\documentclass[latin9]{aa}
\usepackage[T2A,T1]{fontenc}

\usepackage{array}
\usepackage{booktabs}
\usepackage{multirow}
\usepackage{amstext}
\usepackage{amssymb}
\usepackage{wasysym}
\usepackage{graphicx}
\usepackage{txfonts}
\usepackage{setspace}
\usepackage{esint}

\providecommand{\tabularnewline}{\\}

\newcommand{\lyxdot}{.}

\begin{document}

\title{Self-consistent Monte Carlo simulations of proton acceleration in
coronal shocks: Effect of anisotropic pitch-angle scattering of particles}

\author{A. Afanasiev, M. Battarbee, and R. Vainio }

\institute{Department of Physics and Astronomy, University of Turku, Turku,
Finland\\
     \email{alexandr.afanasiev@utu.fi}
     }

 \date{Received ; accepted }
 
\titlerunning{Self-consistent Monte Carlo simulations of proton acceleration}
\authorrunning{Afanasiev et al.}

\abstract
{Solar energetic particles observed in association with coronal mass
ejections (CMEs) are produced by the CME-driven shock waves. The acceleration
of particles is considered to be due to diffusive shock acceleration
(DSA).}
{We aim at a better understanding of  DSA in the case of quasi-parallel
shocks, in which self-generated turbulence in the shock vicinity plays
a key role. }
{We have developed and applied a new Monte Carlo simulation code for
acceleration of protons in parallel coronal shocks. The code performs
a self-consistent calculation of resonant interactions of particles
with Alfv\'en waves based on the quasi-linear theory. In contrast
to the existing Monte Carlo codes of DSA, the new code features the
full quasi-linear resonance condition of particle pitch-angle scattering.
This allows us to take anisotropy of particle pitch-angle scattering 
into account, while the older codes implement an approximate resonance
condition leading to isotropic scattering. We  performed simulations
with the new code and with an old code, applying the same initial and
boundary conditions, and have compared the results provided by both
codes with each other, and with the predictions of the steady-state
theory. }
{We have found that anisotropic pitch-angle scattering leads to less
efficient acceleration of particles than isotropic. However, extrapolations
to particle injection rates higher than those we were able to use
suggest the capability of DSA to produce relativistic particles. The
particle and wave distributions in the foreshock as well as their
time evolution, provided by our new simulation code, are significantly
different from the previous results and from the steady-state theory.
Specifically, the mean free path in the simulations with the new code
is increasing with energy, in contrast to the theoretical result.}
{}


\maketitle

\section{Introduction}

Many eruptive phenomena on the Sun, such as solar flares and
coronal mass ejections (CMEs), are accompanied by outbursts
of energetic charged particles known as solar energetic particle
(SEP) events. The events associated with CMEs are thought to be produced
by shock waves driven by  ejections moving rapidly through solar wind plasma. 
The main acceleration mechanism of ions at these kinds of
shocks is considered to be diffusive shock acceleration (DSA, \citealt{AxfordLeerSkadron-1977};
\citealt{Krymskii-1977}; \citealt{BlandfordOstriker-1978}; \citealt{Bell-1978}),
in which particles get energized by repeatedly crossing the shock.
In the case of quasi-parallel shocks, an essential role in the mechanism
is given to the Alfv\'enic turbulence in the vicinity of the shock,
which scatters particles in pitch angle and facilitates the return
of particles back to the shock. On the other hand, efficient operation
of DSA requires the turbulence ahead of the shock, i.e. in the foreshock,
to be much stronger than that expected in the ambient corona/interplanetary
space. This is achieved by means of the accelerated particles themselves,
which generate Alfv\'en waves due to a streaming instability (\citealt{Bell-1978}).
This couples the processes of particle acceleration and turbulence
evolution.

There are a number of analytical theories of particle acceleration
at shocks, accounting for the particle-wave coupling (\citealt{Bell-1978};
\citealt{Lee-1983}, \citeyear{Lee-2005}; \citealt{GordonLeeMobius-1999}),
which, however, rely on the assumption of a steady state for particles
and waves. In other words, even if the shock parameters change with
time, e.g. because of the shock propagation away from the Sun, the particle/wave
distributions represent a temporal sequence of stationary solutions
depending on the current shock parameters. In spite of the progress
associated with the proposed theories, there has been an interest
to take the dynamical behaviour of the system into
account. As a result of the complexity of the problem, this has been done
with numerical simulations. One of the approaches was suggested by
\citet{ZankRiceWu-2000} and further developed in, e.g. \citeauthor{LiZankRice-2003}
(\citeyear{LiZankRice-2003}, \citeyear{LiZankRice-2005}). With this
approach, the transport of particles in the shock vicinity is treated
numerically, but the accelerated particle spectrum at the shock is
approximated analytically. Further development has been pointed towards
a completely self-consistent dynamic modelling of particle acceleration
(for the results, see e.g. \citealt{VainioLaitinen-2007}, \citeyear{VainioLaitinen-2008};
\citealt{NgReames-2008}; \citealt{BattarbeeLaitinenVainio-2011};
\citealt{Battarbee-2013}; \citealt{VainioPonniBattarbee-2014}). 

Self-consistent Monte Carlo simulations are one of the approaches
in constructing dynamical models of particle acceleration. Since these simulations are based
on the consideration of individual particles interacting with turbulence,
they can provide detailed information on particle and
turbulence distributions in the vicinity of the shock. Current Monte
Carlo simulations use the quasi-linear theory (\citealt{Jokipii-1966})
applicable for interactions of particles with weak slab-mode turbulence
(represented by e.g. the spectrum of low-amplitude Alfv\'en waves).
In  quasi-linear theory, wave-particle interactions are resonant
and governed by the following resonance condition (written in the
wave rest frame):

\begin{equation}
k_{\mathrm{res}}=\frac{\Omega}{\upsilon\mu},\label{eq: full res condition}
\end{equation}
where $k_{\mathrm{res}}$ is the (resonant) wavenumber, $\upsilon$
and $\mu$ are the particle speed and pitch-angle cosine as measured
in the wave rest frame, $\Omega=\gamma^{-1}\Omega_{0}$, $\Omega_{0}$
is the particle cyclotron frequency and $\gamma$ is the relativistic
factor. Furthermore, particle pitch-angle scattering is determined
by the pitch-angle diffusion coefficient
\begin{equation}
D_{\mu\mu}(\mu)=\frac{\pi}{2}\Omega\frac{|k_{\mathrm{res}}|I_{\mathrm{w,res}}}{B^{2}}(1-\mu^{2}),\label{eq: anisotrop diff coef}
\end{equation}
where $B$ is the mean magnetic field and $I_{\mathrm{w,res}}$ is
the value of the wave intensity spectrum $I_{\mathrm{w}}(k)$ at $k=k_{\mathrm{res}}$. 

In recent years, the Coronal Shock Acceleration (CSA) Monte Carlo
code (\citealt{VainioLaitinen-2007}, \citeyear{VainioLaitinen-2008};
\citealt{BattarbeeLaitinenVainio-2011}; \citealt{Battarbee-2013})
has been extensively used to study particle acceleration and foreshock
evolution in coronal shocks. The advantage of the code is that it
allows for simulations of acceleration in
parallel or oblique shocks on a global spatial scale (the shock can
cover a distance of tens of solar radii in a single simulation). However,
its performance is largely achieved via a simplified pitch-angle
resonance condition, 
\begin{equation}
k_{\mathrm{res}}=\frac{\Omega}{\upsilon},\label{eq:simplified res condition}
\end{equation}
which leads to isotropic pitch-angle scattering
\begin{equation}
D_{\mu\mu}(\mu)=\frac{1}{2}\nu(1-\mu^{2}),\label{eq: Isotrop diff coef}
\end{equation}
where the particle scattering rate $\nu=\pi\Omega^{2}I_{\mathrm{w,res}}/(\upsilon B^{2})$
does not depend on the pitch angle. The simplified resonance condition
implies that particles at a particular speed interact with only one
particular spectral component of the wave intensity spectrum. Moreover,
Eq. (\ref{eq: Isotrop diff coef}) shows that the resonance gap does
not appear (instead, the pitch-angle diffusion coefficient has a maximum
at $\mu=0$). Although it is generally accepted that quasi-linear
theory does not work close to $\mu=0$ (see e.g. introduction in
\citealt{NgReames-1995} and references therein) and particle acceleration
and transport simulation studies usually employ so-called filling of the
resonance gap, one can suspect an over-efficiency of particle acceleration
in CSA simulations. Also, the ability of particles to interact with
multiple wave spectral components creates new effects that are not
reproduced if the simplified pitch-angle scattering model is used
(\citealt{NgReames-2008}).

To study the effect of anisotropic pitch-angle scattering on particles
acceleration at coronal shocks, we developed a new Monte Carlo
code named SOLar Particle Acceleration in Coronal Shocks (SOLPACS).
Contrary to the CSA code, this code utilizes the full quasi-linear
resonance condition of pitch-angle scattering (\citealt{AfanasievVainio-2013}).
In the present paper, we compare SOLPACS simulation results with those
of CSA, obtained for the same initial and boundary conditions, and
with the steady-state theory of diffusive shock acceleration (\citealt{Bell-1978},
see also \citealt{VainioLaitinen-2007}, \citeyear{VainioLaitinen-2008}
and \citealt{VainioPonniBattarbee-2014}). 

Further, in Sect. 2, we review some aspects of the steady-state
theory, which are further used in the analysis of the simulations.
In Sect. 3, we describe the simulation set-up. In Sect. 4, we present
simulation results and their comparative analysis. In Sect. 5, we
provide further discussion of SOLPACS and CSA results. Finally, in
Sect. 6, we present our conclusions.

\section{Steady-state theory}

Here we specify the accelerated particles by the conventional distribution
function $f=d^{6}N/(dx^{3}\, dp^{3})$ in phase space, and Alfv\'en
waves by the wave intensity $I_{\mathrm{w}}$, which is defined by
\begin{equation}
\delta B^{2}=\intop_{-\infty}^{+\infty}I_{\mathrm{w}}(k)dk,\label{eq: Definition of wave intensity spectrum}
\end{equation}
where the positive and negative values of $k$ correspond to the waves
with opposite circular polarizations. Then, Bell's steady-state theory
of diffusive shock acceleration at a (planar) parallel shock can be
presented in the following form (\citealt{VainioLaitinen-2007}; \citealt{VainioPonniBattarbee-2014}):
\begin{equation}
f(x,p)=f_{\mathrm{s}}\frac{x_{0}}{x+x_{0}},\label{eq: SST part distrib func}
\end{equation}
\begin{equation}
I_{\mathrm{w}}(x,k)=\frac{1}{3\pi}\frac{\Omega B^{2}|k|^{-3}}{(u_{1}-V_{\mathrm{A}})(x+x_{0})},\label{eq:SST wave intensity}
\end{equation}
where

\begin{equation}
f_{\mathrm{s}}=f(0,p)=\frac{\sigma\epsilon_{\mathrm{inj}}n_{\mathrm{0}}}{4\pi p_{\mathrm{inj}}^{3}}\left(\frac{p}{p_{\mathrm{inj}}}\right)^{-\sigma},\label{eq: SST shock distr func}
\end{equation}
\begin{equation}
x_{0}(p)=\frac{2}{\pi\epsilon_{\mathrm{inj}}\sigma}\frac{V_{\mathrm{A}}}{\Omega_{\mathrm{0}}}\left(\frac{p}{p_{\mathrm{inj}}}\right)^{\sigma-3},\label{eq: SST x0}
\end{equation}
$x$ is the distance from the shock towards the upstream region along
the mean magnetic field; $u_{1}$ is the shock-frame solar wind speed;
$V_{\mathrm{A}}$ is the Alfv\'en speed; $n_{\mathrm{0}}$ is the
thermal proton density in the shock's upstream; $\sigma=3r_{\mathrm{c}}/(r_{\mathrm{c}}-1)$
is the spectral index of the particle distribution function at shock;
$r_{\mathrm{c}}=V_{1}/V_{2}$ is the scattering-centre compression
ratio, which is defined as the ratio of the shock-frame phase velocities
$V_{1}$ and $V_{2}$ of Alfv\'en wave modes in the upstream and
downstream regions, respectively (see, \citealt{VainioSchlickeiser-1998}
for details); and $\epsilon_{\mathrm{inj}}$ is the fraction of upstream
protons injected into the acceleration process at the injection momentum,
$p_{\mathrm{inj}}$. The parameter $x_{0}(p)$ in Eq. (\ref{eq: SST x0})
has to be expressed as a function of $k$, before it can be substituted
to Eq. (\ref{eq:SST wave intensity}), applying the resonance condition,
$p=m_{\mathrm{p}}\Omega_{0}/|k|$. The parameter $\epsilon_{\mathrm{inj}}$
is related to the particle injection rate $Q$ (the number of particles
injected at the shock per unit cross-section of the shock and unit
time) through $Q=\epsilon_{\mathrm{inj}}n_{\mathrm{0}}u_{1}$, where
$n_{0}$ is the thermal proton density in the shock's upstream. The
parameter $p_{\mathrm{inj}}$ can be calculated from (\citealt{VainioLaitinen-2007})
\begin{equation}
p_{\mathrm{inj}}^{\sigma-3}=\frac{1}{N_{\mathrm{inj}}}\intop_{0}^{\infty}\frac{dN_{\mathrm{inj}}}{dp}p^{\sigma-3}dp,\label{eq: Effective injection momentum}
\end{equation}
where $dN_{\mathrm{inj}}/dp$ is the injected particle spectrum.

Using Eqs. (\ref{eq: SST part distrib func}) and (\ref{eq: SST shock distr func}),
one can calculate the omni-directional particle intensity $I(x,p)=p^{2}f(x,p)$
and, using energy $E$ as an independent variable, obtain at the shock
(in the non-relativistic regime): $I(E)\propto E^{1-\sigma/2}$. Considering
the wave intensity $I_{\mathrm{w}}(x,k)$, it is easy to see from
Eq. (\ref{eq: SST x0}) that $x_{0}(k)\propto|k|^{3-\sigma}$ ($\sigma>3$).
At a given distance $x$ from the shock, defining $k_{\mathrm{b}}$
as the solution of $x_{0}(k)=x$, Eq. (\ref{eq:SST wave intensity})
reveals that the wave intensity represents a broken power law in $k$:
at $|k|\gg k_{\mathrm{b}}$, we obtain $I_{\mathrm{w}}(k)\propto|k|^{-3}$,
and at $|k|\ll k_{\mathrm{b}}$, we get $I_{\mathrm{w}}(k)\propto|k|^{\sigma-6}$.
The break point $k_{\mathrm{b}}$ decreases with distance from the
shock. 

Additionally, one can calculate the particle mean free path $\lambda$,
which is defined as

\begin{equation}
\lambda=\frac{3\upsilon}{8}\intop_{-1}^{+1}\frac{(1-\mu^{2})^{2}}{D_{\mu\mu}}d\mu,\label{eq:mfp_definition}
\end{equation}
 but in Bell's steady-state DSA theory, after applying the simplified
resonance condition, it reads (\citealt{VainioPonniBattarbee-2014})
\begin{equation}
\lambda(x,\upsilon)=\frac{3(u_{1}-V_{\mathrm{A}})}{\upsilon}(x+x_{0}).\label{eq:SST mfp}
\end{equation}

Finally, one can estimate the cut-off energy in the particle energy
spectrum, $I(E)$, which can occur because of, e.g. a finite acceleration
time, using the mean free path resulting from Bell's steady-state
theory. In that case, assuming a parallel shock and constant parameters
specifying the shock and  plasma (which is the case in our simulations),
one can obtain an estimate for the cut-off momentum $p_{\mathrm{c}}$
for the shock propagating in time $t_{\mathrm{s}}$ (\citealt{VainioPonniBattarbee-2014}), i.e.
\begin{equation}
p_{\mathrm{c}}^{\sigma-3}=p_{\mathrm{inj}}^{\sigma-3}\left(1+\frac{\pi}{2}\frac{\epsilon_{\mathrm{inj}}\sigma(M_{\mathrm{A}}-1)}{r_{\mathrm{c}}}\Omega_{\mathrm{0}}t_{\mathrm{s}}\right),\label{eq: Cut-off momentum}
\end{equation}
where $M_{\mathrm{A}}=u_{1}/V_{\mathrm{A}}$ is the Alv\'enic Mach
number of the upstream flow.

\section{Simulation set-up}

In both CSA and SOLPACS, particles are traced within a 1-D spatial
simulation box, i.e. along the mean magnetic field line. Although
CSA allows variation of the plasma density and magnetic field with
distance from the Sun, here we focus on effects produced by the two
different wave-particle interaction models. Therefore, we perform
simulations for a simple set-up, assuming constant thermal proton density,
$n_{0}=3.6\times10^{12}\,\mathrm{m^{-3}}$, magnetic field, $B_{0}=3.4\times10^{-5}\,\mathrm{T}$
and solar wind speed, $u_{0}=12.4\:\mathrm{km\, s^{-1}}$ in the whole
simulation domain. The applied values correspond to the density, magnetic
field, and solar wind speed given by the coronal model of \citet{VainioLaitinen-2007}
for the heliocentric distance of $2R_{\odot}$. The implied locality
of the simulations allows us to consider the simulation box to be
moving with the shock, i.e. in the shock frame. In this case, simulating
only the upstream region, we take the shock itself as one boundary
of the box. The other boundary is taken to be at a distance of $1R_{\odot}$
from the shock. Further, we assume the plasma turbulence in the box
to be due to outwards-propagating (if considered in the solar wind
frame) Alfv{\'e}n waves, with the initial spectral form $\propto k^{-3/2}$.
We select a low initial wave intensity, such that it provides
little pitch-angle scattering for injected particles, i.e. a non-diffusive
particle transport, in the spatial simulation box at early times in
a simulation. Specifically, it is calculated assuming the initial
mean free path $\lambda=1R_{\odot}$ for 100 keV protons in both CSA
and SOLPACS%
\footnote{The calculations are performed using Eq. (\ref{eq:mfp_definition})
and the initial power-law wave spectrum. No resonance gap filling
(see text) is taken into account.%
}. We simulate protons, which are assumed to be injected at the shock
at a constant rate $Q$, which is characterized by the injection efficiency
$\epsilon_{\mathrm{inj}}$. Similar to \citet{VainioLaitinen-2007},
the injected particles are characterized by an exponential velocity
spectrum (in the shock frame) $dN_{\mathrm{inj}}/d\upsilon=(N_{\mathrm{inj}}/\upsilon_{1})H(\upsilon-u_{1})\mathrm{e^{-(\upsilon-u_{1})/\upsilon_{1}}}$,
where $N_{\mathrm{inj}}=Qt$ is the number of injected particles per
unit cross-section of the magnetic flux tube injected in time $t$,
$\upsilon_{1}$ is a parameter, and $H$ is the Heaviside step function.
Similar to \citet{VainioLaitinen-2007}, instead of tracing particles
in the shock's downstream, we employ a probability of return from
the downstream (\citealt{JonesEllison-1991}), $P_{\mathrm{ret}}=(\upsilon_{\mathrm{w}}-V_{2})^{2}/(\upsilon_{\mathrm{w}}+V_{2})^{2}$,
where $\upsilon_{\mathrm{w}}$ is the particle speed in the frame
of the downstream waves and $V_{2}=(u_{1}-V_{\mathrm{A}})/r_{\mathrm{c}}$
is the effective speed of dowstream Alfv\'en waves as measured in
the shock frame. In our simulations, we take the shock speed $V_{\mathrm{s}}=1500\,\mathrm{km\, s^{-1}}$
and fix the scattering-centre compression ratio to $r_{\mathrm{c}}=4$
and $\upsilon_{1}=375\,\mathrm{km\, s^{-1}}$. For the assumed values,
from Eq. (\ref{eq: Effective injection momentum}) we obtain $p_{\mathrm{inj}}=m_{\mathrm{p}}(u_{1}+\upsilon_{1})$,
which gives $E_{\mathrm{inj}}=p_{\mathrm{inj}}^{2}/(2m_{\mathrm{p}})=18.1\,\mathrm{keV}$.

The numerical scheme to simulate wave-particle interactions, implemented
in SOLPACS, is detailed in \citet{AfanasievVainio-2013}. In the following,
we  recall two main points important for the analysis of simulation
data.

At each time step in a simulation, the wave intensity spectrum $I_{\mathrm{w}}(k)$
is given by its values at the $k$-grid nodes: $\{k_{j}\}$ and by
a set of spectral indices $q$: $\{q_{j}\},$ resulting from interpolation
of those values by a power-law function between every two adjacent
$k$-grid nodes: $q_{j}=\ln(I_{\mathrm{w},j+1}/I_{\mathrm{w},j})/\ln(k_{j+1}/k_{j})$.
Therefore, for $k\in(k_{j},k_{j+1})$, we have $I_{\mathrm{w}}(k)=I_{\mathrm{w},j}(k/k_{j})^{-q_{j}}$.
Considering the $\mu$-grid: $\{\mu_{j}=\Omega/(\upsilon k_{j})\}$,
for $\mu\in(\mu_{j+1},\mu_{j})$, we obtain $\nu(\mu)=\nu_{0,j}|\mu|^{q_{j}-1}$,
where $\nu_{0,j}=\pi\Omega/B^{2}(\upsilon/\Omega)^{q_{j}-1}I_{\mathrm{w},j}|k_{j}|^{q_{j}}$.
Using this notation, the mean free path (Eq. \ref{eq:mfp_definition})
can be calculated as
\begin{equation}
\lambda=\frac{3\upsilon}{4}\sum_{j}\int_{\mu_{j+1}}^{\mu_{j}}\frac{1-\mu^{2}}{\nu_{0,j}|\mu|^{q_{j}-1}}d\mu,\label{eq:mfp-numerical calculation}
\end{equation}
where summation is performed over the grid nodes in $\mu$-space.

The employed pitch-angle scattering method requires extremely small
time steps to be taken near $\mu=0$. Therefore, we fill the
resonance gap by approximating the scattering frequency $\nu(\mu)$
near $\mu=0$ by a linear function $\widetilde{\nu}(\mu)=b\mu+b_{1}$
in the interval $(-\mathrm{\mu_{min},\mu_{\mathrm{min}})}$, where
$\mu_{\mathrm{min}}=\Omega/(\upsilon k_{\mathrm{max}})$ and $k_{\mathrm{max}}$
is the maximum node of the $k$-grid, which is taken to be about $\Omega_{0}/V_{\mathrm{A}}$.
In these simulations, $k_{\mathrm{max}}=10^{-2}\,\mathrm{m^{-1}}$.
The technique applied makes the resonance gap in the $\mu$-space
narrower for higher-energy particles than for low-energy particles. 

\begin{figure*}[t!]
\begin{centering}
\includegraphics[width=\columnwidth]{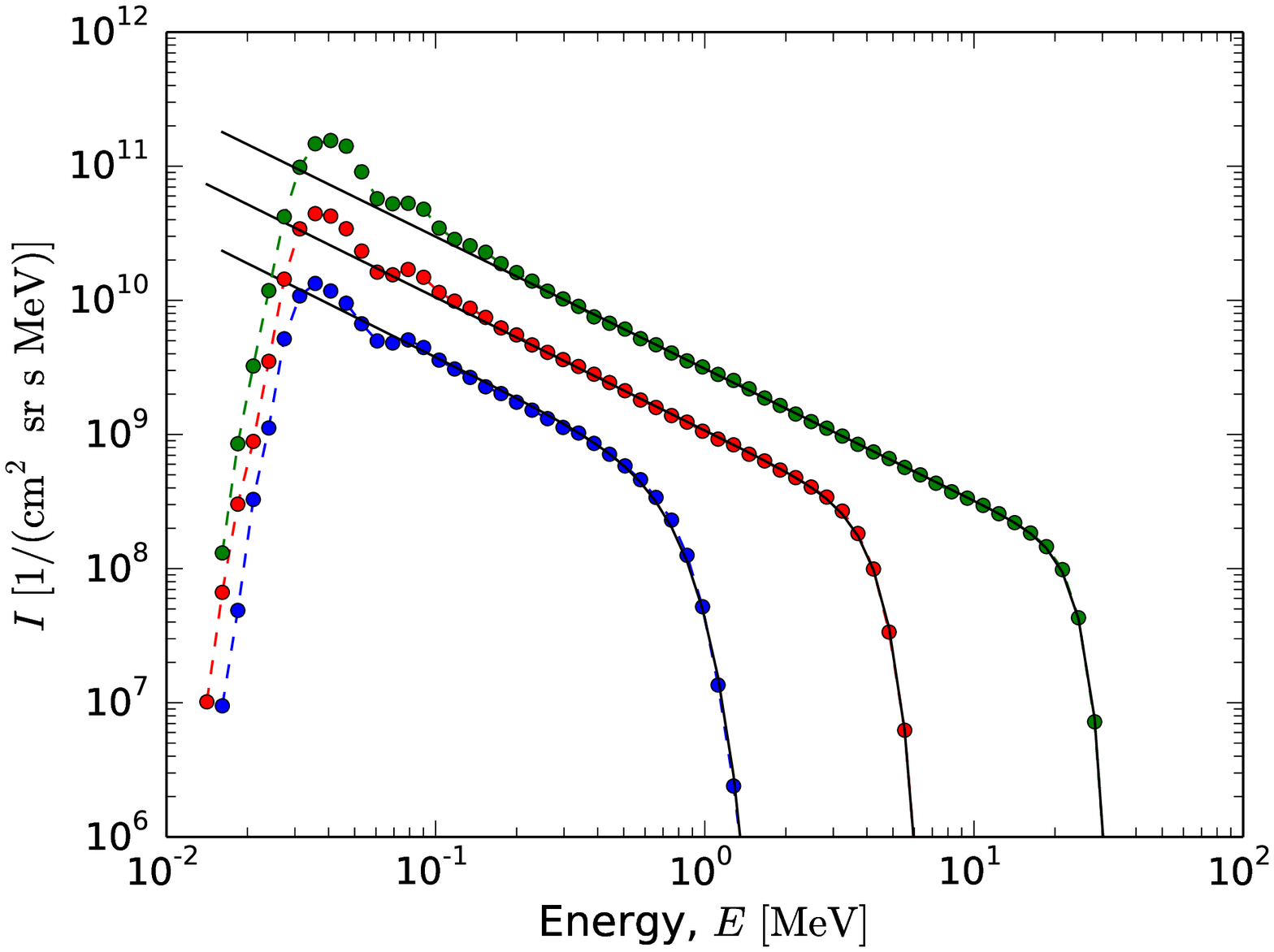}\includegraphics[width=\columnwidth]{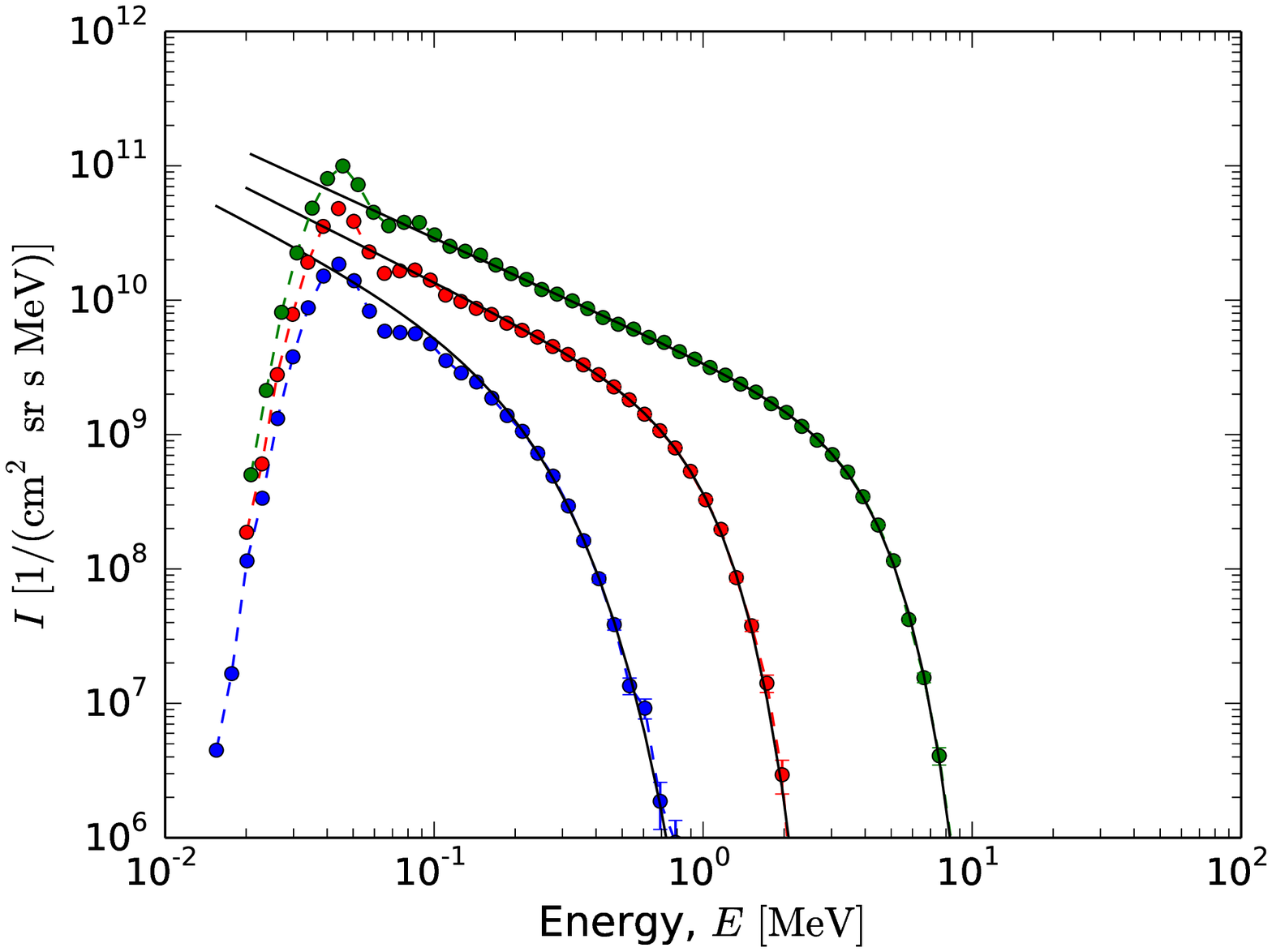}
\par\end{centering}

\protect\caption{\label{fig:Simulated energy spectra}Simulated particle energy spectra obtained
with CSA (left panel) and SOLPACS (right panel) at $t=580\,\mathrm{s}$
for different values of the injection efficiency: $\epsilon_{\mathrm{inj}}=1.62\times10^{-6}$
(blue circles), $\epsilon_{\mathrm{inj}}=5.40\times10^{-6}$ (red
circles), and $\epsilon_{\mathrm{inj}}=1.62\times10^{-5}$ (green circles). Black curves represent fits to the spectral data,
obtained with Eq. (\ref{eq:spectrum fitting function}). The fitting was carried out at energies $E>0.2\,\mathrm{MeV}$ in
all cases. }
\end{figure*}

\begin{table*}[t!]
\protect\caption{Best-fit values of the model function (Eq. \ref{eq:spectrum fitting function})
parameters with the corresponding standard deviation errors (given
in parentheses). The pfu unit stands for the particle flux unit ($1\,\mathrm{pfu=1\, cm^{-2}sr^{-1}s^{-1}MeV^{-1}}$).\label{tab:Best-fit-parameters}}

\centering{}
\begin{tabular}{l>{\raggedright}p{1.5cm}>{\raggedright}p{2.75cm}>{\raggedright}p{2.75cm}>{\raggedright}p{2.75cm}>{\raggedright}p{2.75cm}}
\toprule 
\multirow{2}{2.0cm}{Simulation \linebreak type} &  &  & $E_{\mathrm{c}}\,(\sigma_{E_{\mathrm{c}}})$ &  & $C\,(\sigma_{C})\times10^{-9}$\tabularnewline
 & $\epsilon_{\mathrm{inj}}\times10^6$ & $\beta\left(\sigma_{\beta}\right)$ & (MeV) & $\delta\,(\sigma_{\delta})$ & (pfu MeV$^{\beta}$)\tabularnewline
\midrule
\multirow{3}{*}{CSA} & 1.62 & 1 (fixed) & $0.778\,(0.003)$ & $3.09\,(0.02)$ & $0.377\,(0.005)$\tabularnewline
 & $5.40$ & $0.992\,(0.004)$ & $4.27\,(<0.01)$ & $4.79\,(0.01)$ & $1.07\,(<0.01)$\tabularnewline
 & $16.2$ & $0.984\,(0.002)$ & $23.9\,(<0.1)$ & $6.26\,(0.01)$ & $3.10\,(0.01)$\tabularnewline
\midrule
\multirow{3}{*}{SOLPACS} & $1.62$ & 1 (fixed) & $0.180\,(0.034)$ & $1.39\,(0.19)$ & $0.804\,(0.233)$\tabularnewline
 & $5.40$ & $1.00\text{\,}\left(0.08\right)$ & $0.885\,(0.040)$ & $2.17\,(0.11)$ & $1.35\,(0.14)$\tabularnewline
 & $16.2$ & $0.917\,\left(0.010\right)$ & $3.81\,(0.04)$ & $2.35\,(0.04)$ & $3.51\text{\,(}0.03)$\tabularnewline
\bottomrule
\end{tabular}
\end{table*}

\section{Results}

We have carried out simulations for three different values of the
particle injection rate at the shock (given by the parameter $\epsilon_{\mathrm{inj}}$).
The simulations were conducted until the time in a simulation reached
$t_{\mathrm{max}}=580\,\mathrm{s}$, which corresponds to the shock
propagation distance of $1.25\, R_{\odot}$. Figure \ref{fig:Simulated energy spectra}
presents resulting proton energy spectra, $I(E)$, at the shock at
$t=t_{\mathrm{max}}$. The presented spectra are calculated in the
solar-fixed frame (where the shock propagates at speed $V_{\mathrm{s}}$).
Similar to Vainio et al. (2014), we fitted the spectral data by the
function

\begin{equation}
I(E)=CE^{-\beta}\mathrm{exp}\left\{ -\left(\frac{E}{E_{\mathrm{c}}}\right)^{\delta}\right\} ,\label{eq:spectrum fitting function}
\end{equation}
where $C$, $\beta$, $E_{\mathrm{c}}$, and $\delta$ are fitting
parameters. The fits are shown in Fig. \ref{fig:Simulated energy spectra}
as well. The best-fit values of the spectral parameters along with
the errors are given in Table \ref{tab:Best-fit-parameters}. The
indicated errors account for statistical errors in the spectral data
and systematic errors due to binning of the data and the non-power-law
form of the spectra at low energies (visible as quasi-regular variations
of the particle intensity on the plots). In order to avoid large systematic
errors caused by such variations, the spectra were fitted at $E>0.2\,\mathrm{MeV}$.
The details of the fitting algorithm and of the error calculations
are given in Appendix A. When fitting the spectra corresponding
to the weakest injection ($\epsilon{}_{\mathrm{inj}}=1.62\times10^{-6}$),
the power-law index $\beta$ was fixed to 1 (the theoretical value)
as the power-law-like portion of the spectrum to be fitted was quite
small (not recognizable at all in the case of the SOLPACS simulation).
The spectra have similar shapes consisting of a low-energy,
multi-peak component, a power-law part, and a super-exponential tail
in all cases, although the spectra corresponding to anisotropic pitch-angle
scattering have smaller values of the cut-off energy $E_{\mathrm{c}}$
and the cut-off steepness $\delta$. 

\begin{figure*}
\begin{centering}
\includegraphics[width=\columnwidth]{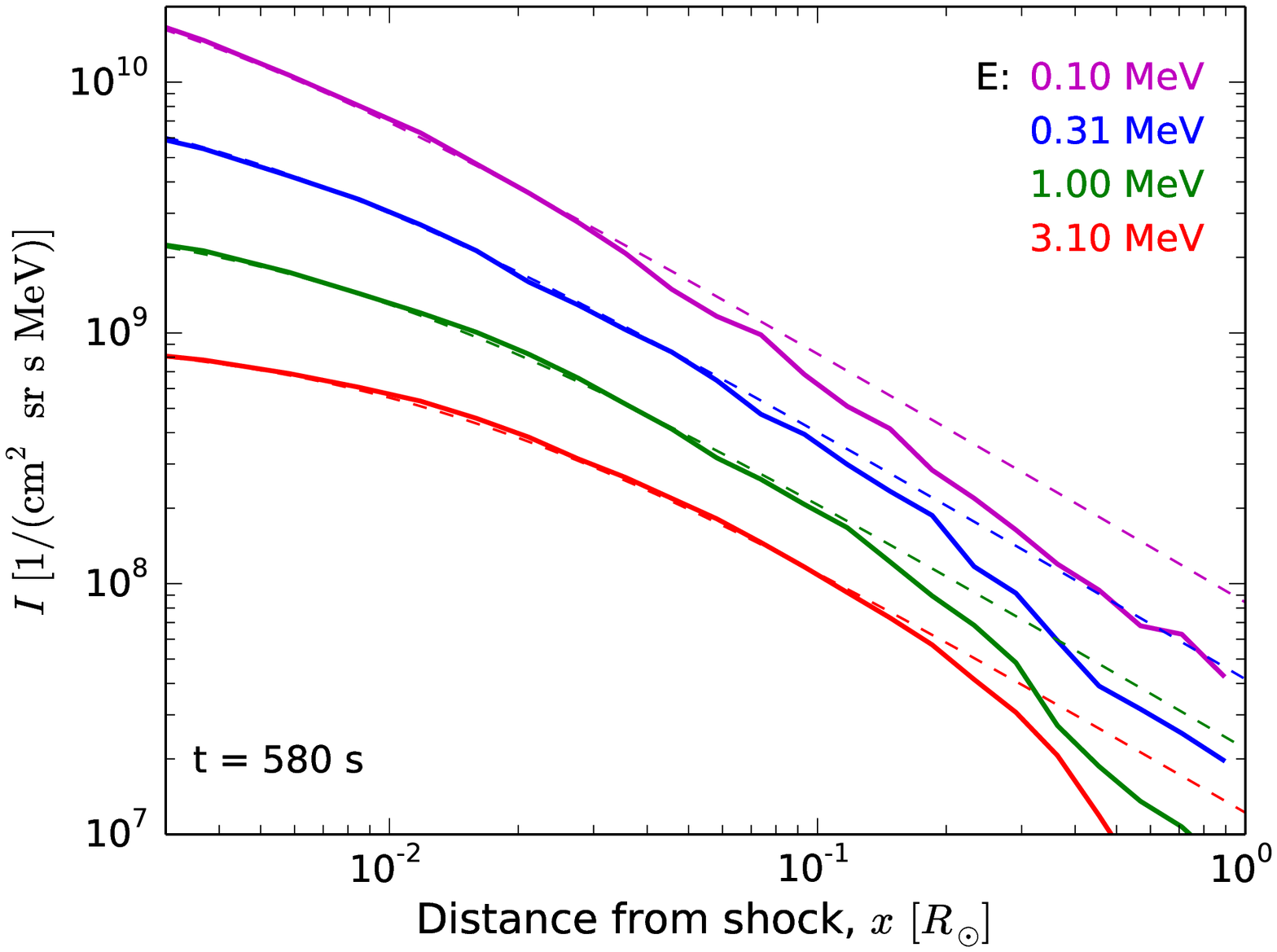}\includegraphics[width=\columnwidth]{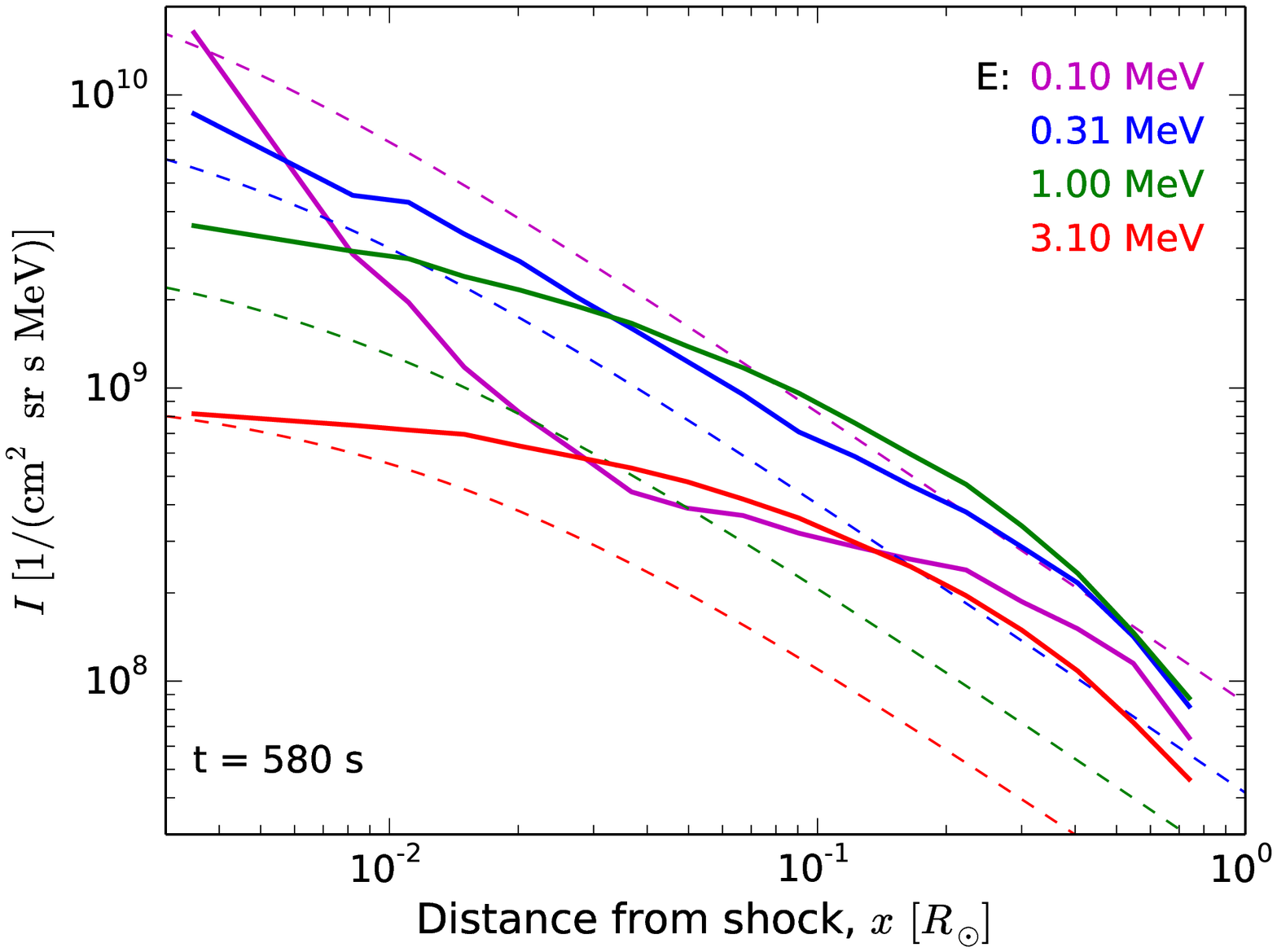}
\par\end{centering}

\protect\caption{\label{fig:Energy distribution in the foreshock}Distribution of particles
in the foreshock at different energies, resulting from CSA (left panel,
continuous curves) and SOLPACS (right panel, continuous curves) at
the end ($t=580\,\mathrm{s}$) of the simulation for $\epsilon_{\mathrm{inj}}=1.62\times10^{-5}$ compared to Bell's steady-state theory (dashed curves). }
\end{figure*}

Figure \ref{fig:Energy distribution in the foreshock} presents spatial
distributions of particles of different energies in the foreshock,
resulting from CSA and SOLPACS in comparison with those resulting
from Bell's steady-state theory. The plotted distributions are calculated
in the mixed reference frame, i.e. the particle intensity $I(x,p')$
represents a function of the distance $x,$ as measured in the shock
frame, and of the momentum $p', $ as measured in the solar fixed frame. 
We apply a correction to the theoretical distributions by multiplying
the injection efficiency $\epsilon_{\mathrm{inj}}$ (Eqs. \ref{eq: SST shock distr func}
and \ref{eq: SST x0}) by a factor $\alpha$, where $1<\alpha<2$,
which accounts for the difference in the pitch-angle distribution
of injected particles in Bell's theory and in our simulations. Bell's
theory injects particles isotropically, whereas our simulations only inject
 into one hemisphere ($\mu>0$) (\citealt{VainioLaitinen-2008}).
We find that $\alpha=1.8$ gives the best correspondence between the
CSA distributions and the theoretical distributions in the vicinity of the
shock. Their deviation, increasing with distance, is due to the free
escape of particles at the outer boundary of the simulation box. In
the case of SOLPACS, the behaviour of particles in the foreshock is
much more complicated than predicted by Bell's steady-state theory. In
 very close vicinity of the shock ($x\lesssim5\times10^{-3}\, R_{\odot}$),
the energy dependence of the particle intensity is similar for both
simulations and theory. However, that
is not the case further in the foreshock. The most dramatic difference is experienced by low-energy
particles: there is a region in the foreshock, in which there is a
lack of these particles in comparison with the theoretical prediction.
And on the contrary, particles of higher energies are in excess compared
to the theory in the whole foreshock region.

\begin{figure*}
\begin{centering}
\includegraphics[width=\columnwidth]{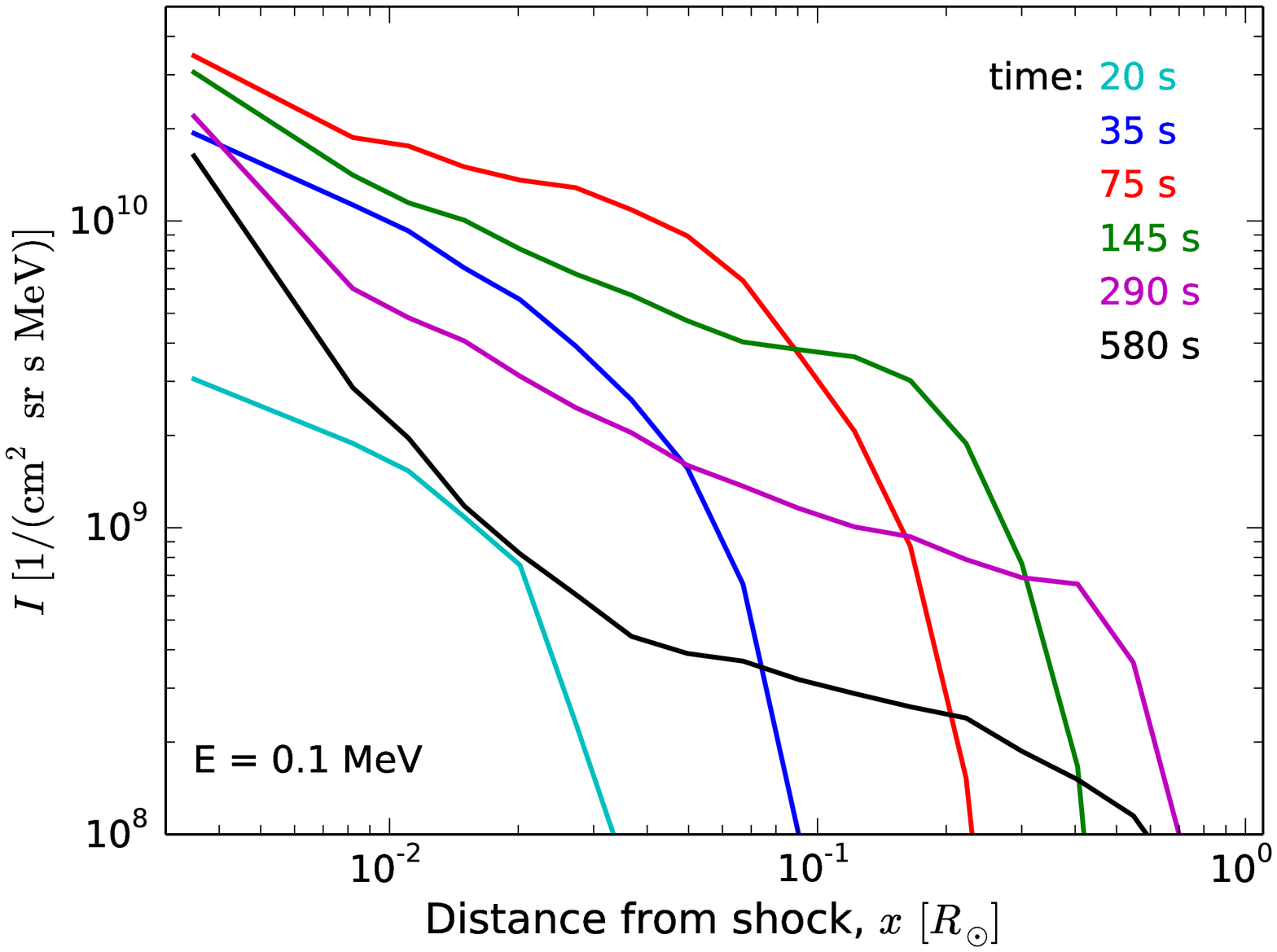}\includegraphics[width=\columnwidth]{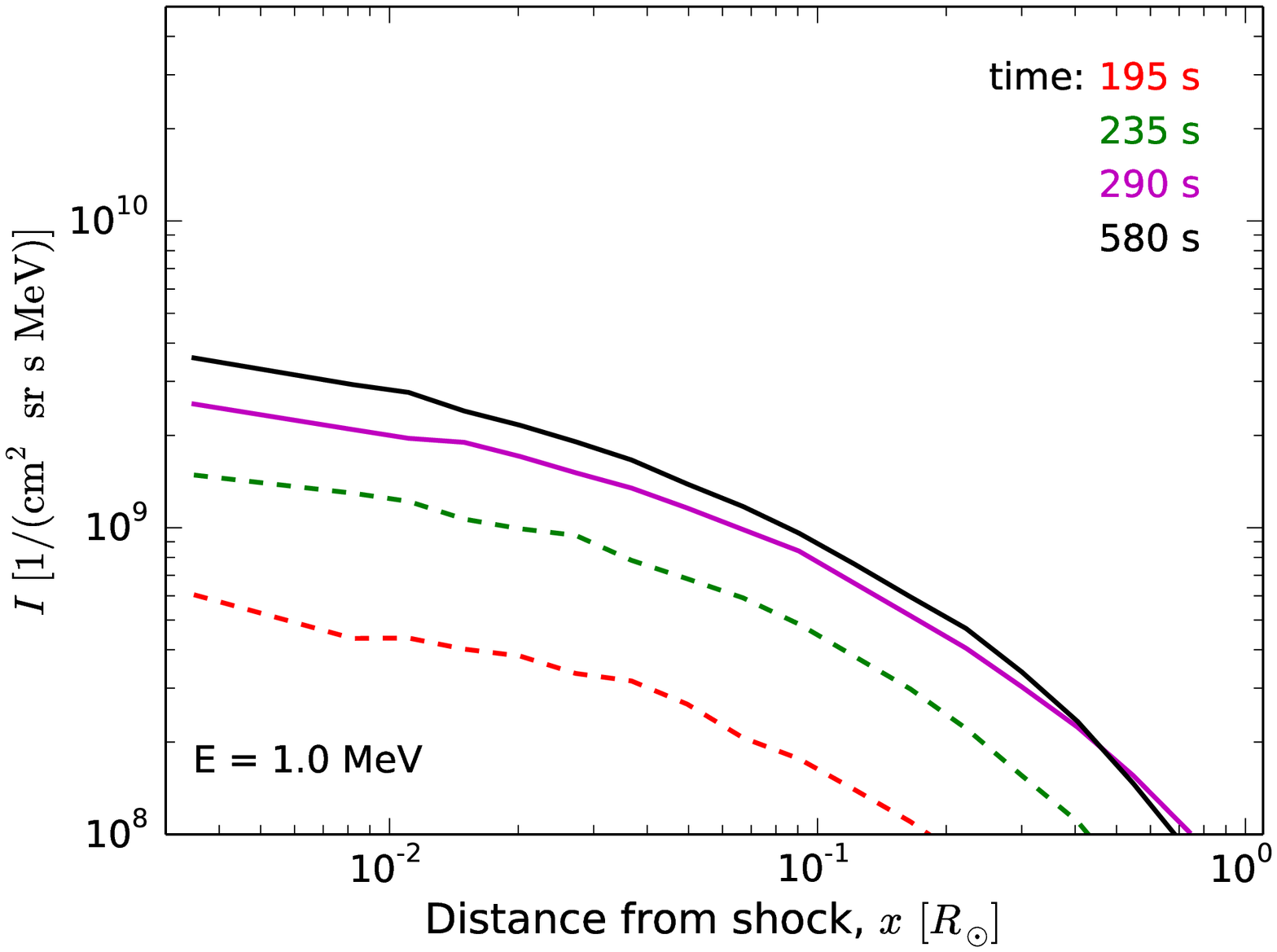}
\par\end{centering}

\protect\caption{\label{fig:Time evolution of spatial distribs}Time evolution of the
spatial distribution of particles in the foreshock at $E=0.1\,\mathrm{MeV}$
(left panel) and $E=1.0\,\mathrm{MeV}$ (right panel) in SOLPACS. Dashed red and green curves in the right panel correspond to additional
times $t=195\,\mathrm{s}$ and $235\:\mathrm{s}$. }
\end{figure*}

To get more information, we looked at the time evolution of spatial
distributions of particles, produced by SOLPACS, at two energies,
0.1 MeV and 1 MeV (Fig. \ref{fig:Time evolution of spatial distribs}).
One can see, in the case of low energy, the particle intensity first
increases with time (at a given location in the foreshock), reaches
a maximum, and then decreases at a much slower rate. In the case of high-energy
particles, the time evolution of their spatial profile is different:
the particle intensity increases with time and largely stagnates close
to its maximum level. There is also a flattening before the steep tail
in the spatial profiles of 0.1 MeV particles. 

\begin{figure*}
\begin{centering}
\includegraphics[width=\columnwidth]{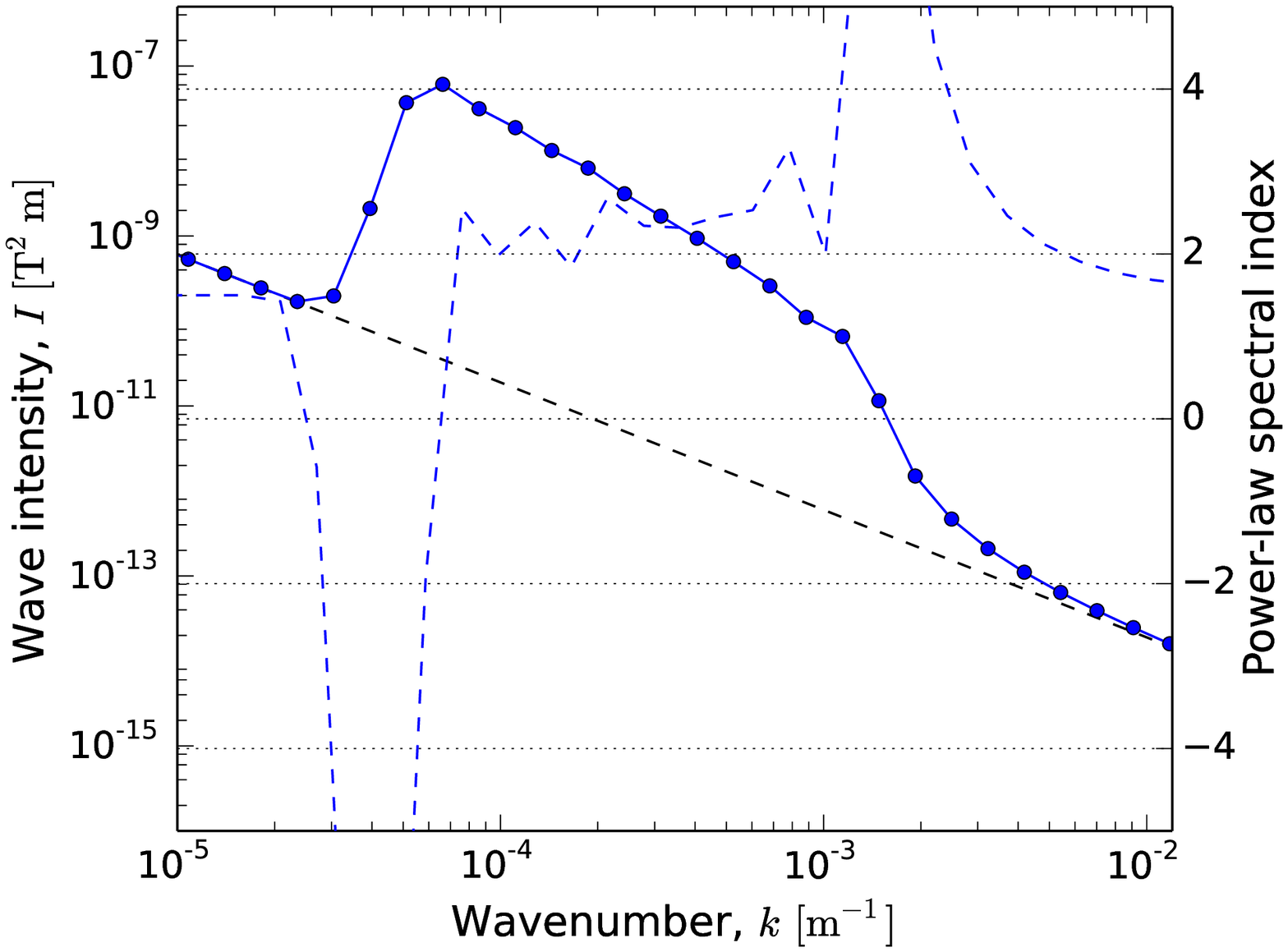}\includegraphics[width=\columnwidth]{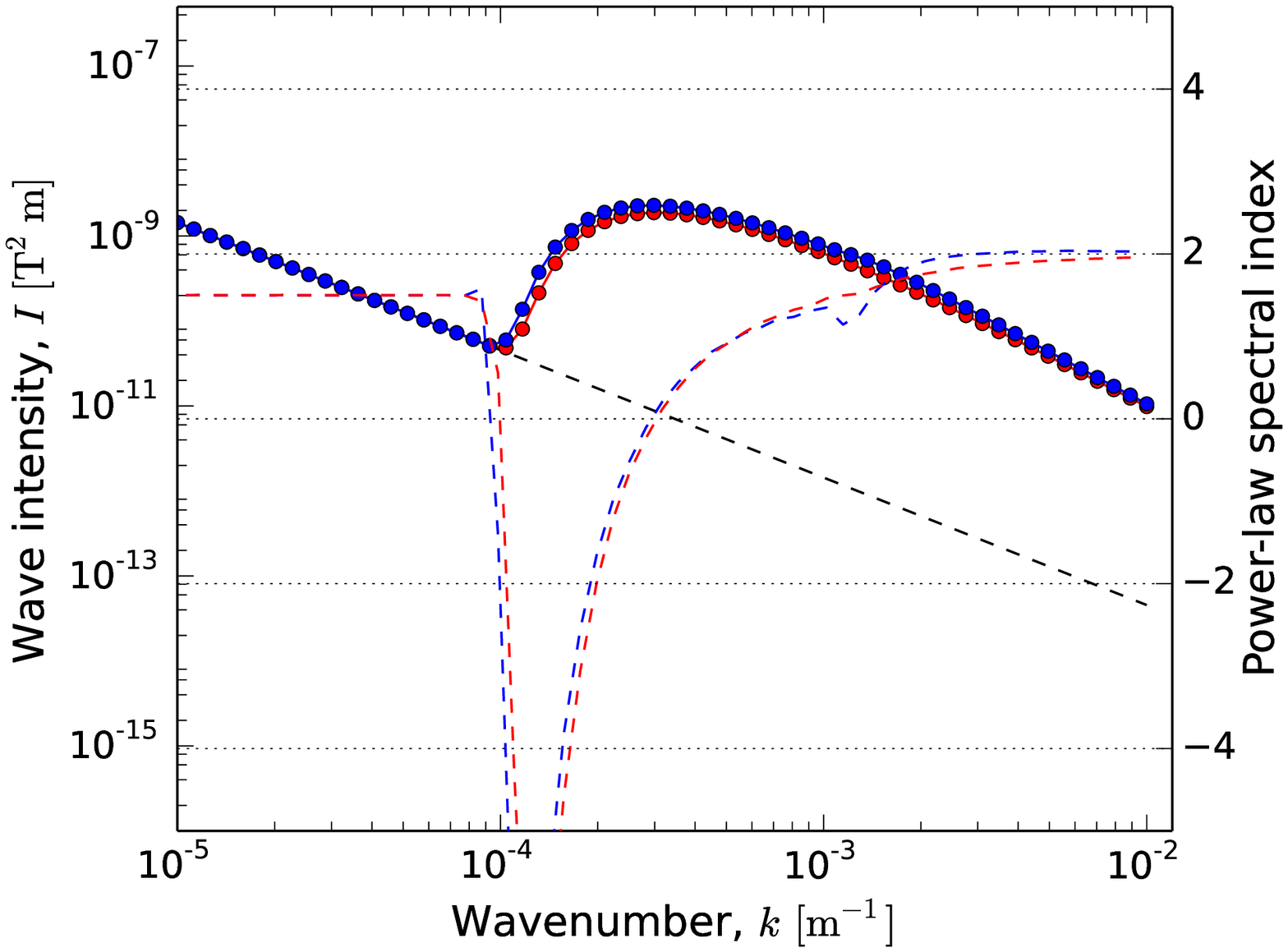}
\par\end{centering}

\begin{centering}
\includegraphics[width=\columnwidth]{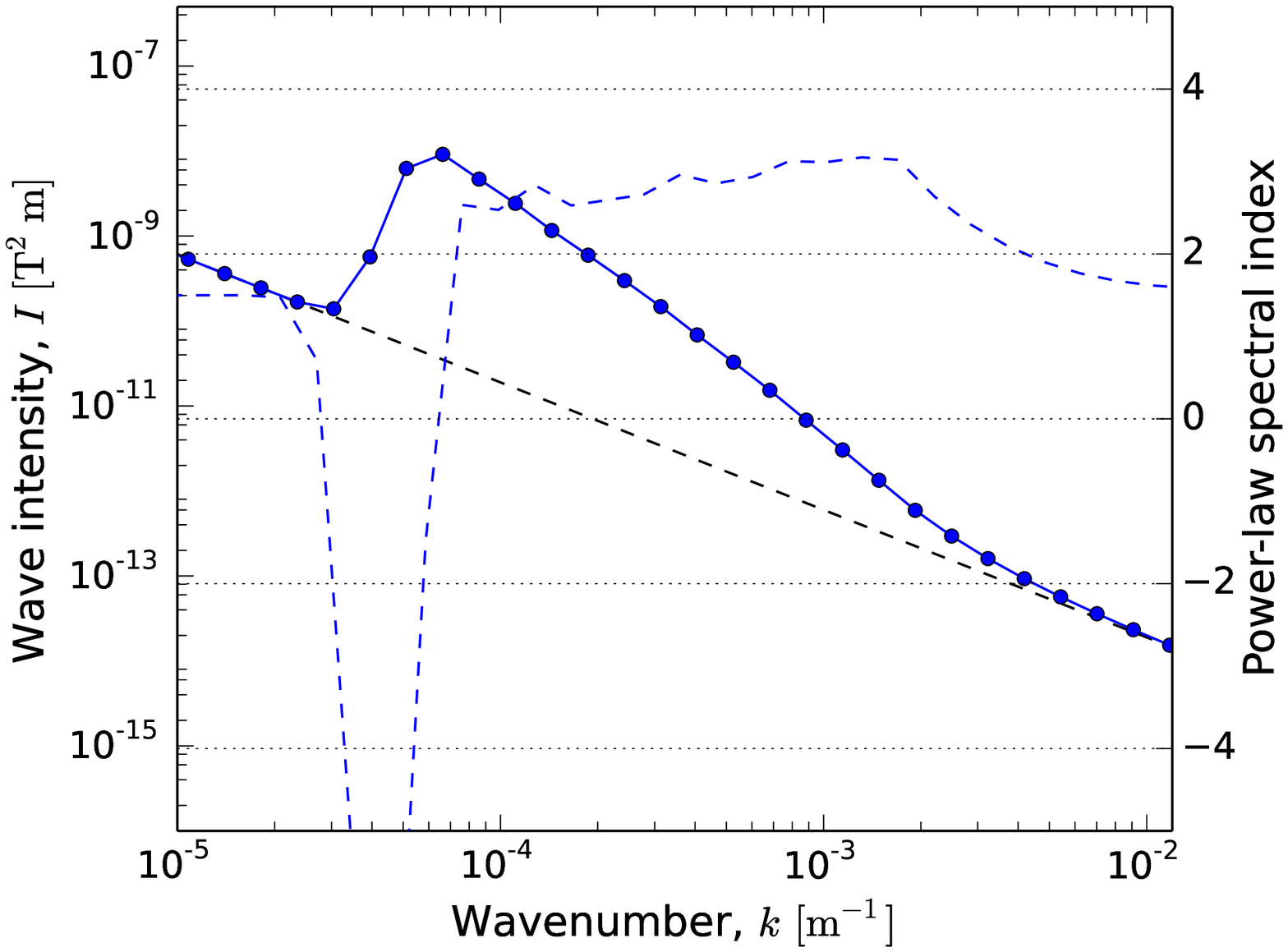}\includegraphics[width=\columnwidth]{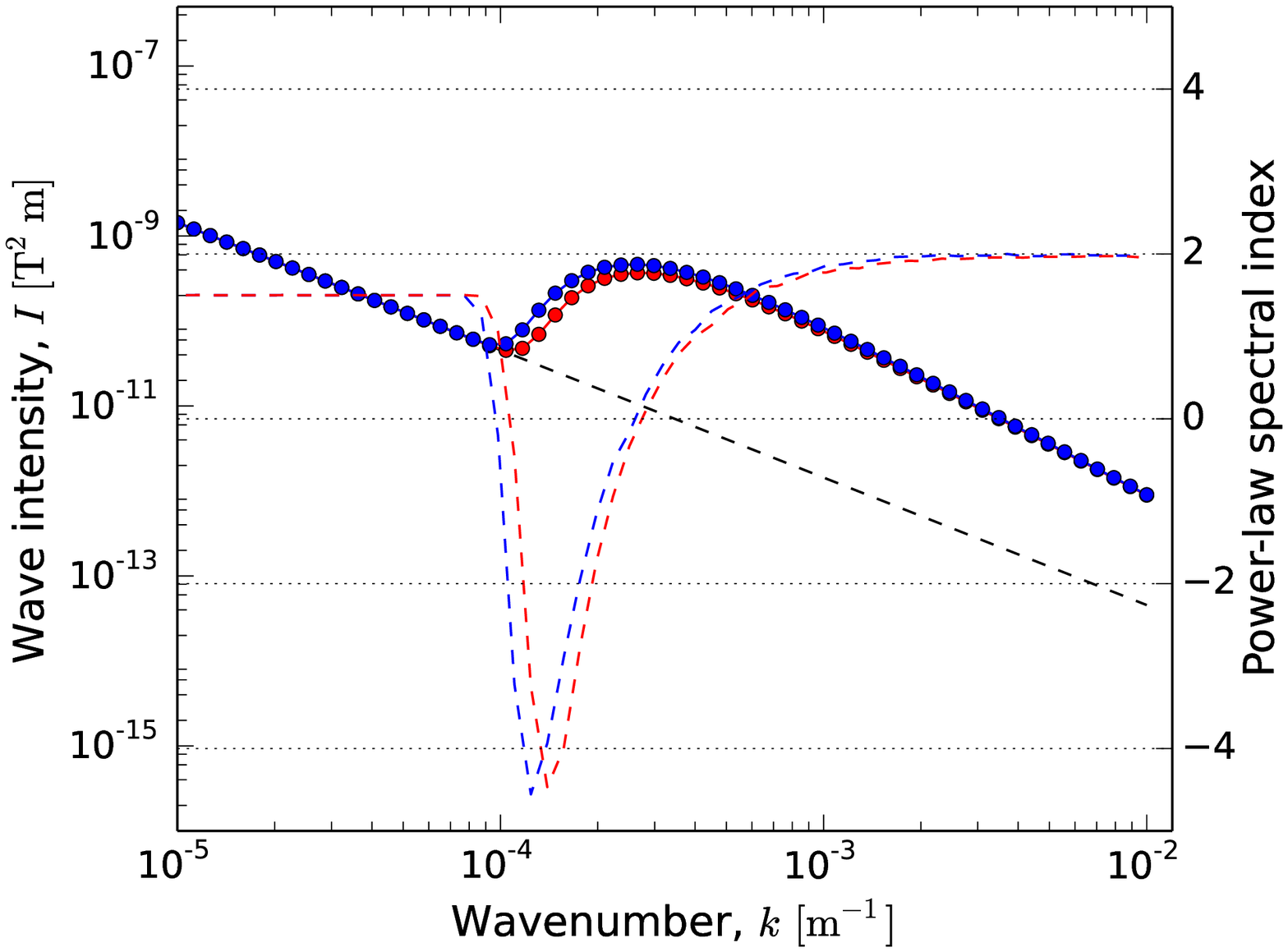}
\par\end{centering}

\protect\caption{\label{fig:Simulated wave spectra}Wave intensity spectra obtained
with CSA (left column) and SOLPACS (right column) for $\epsilon_{\mathrm{inj}}=1.62\times10^{-5}$
at distance $x=0.0035\, R_{\odot}$ (top row) and
$x=0.13\, R_{\odot}$ (bottom row) in front of the shock at $t=580\,\mathrm{s}$.
The filled blue and red circles in the SOLPACS data plots denote
wave intensities of Alfv\'en waves of the opposite circular polarization.
In all  plots,  coloured dashed lines show spectral indices and
 black dashed line shows the initial spectrum, normalized to provide 
$\lambda=1\, R_{\odot}$ for 100 keV protons.}
\end{figure*}

Figure \ref{fig:Simulated wave spectra} shows Alfv\'en wave spectra
at the end ($t=580\,\mathrm{s}$) of the simulations for $\epsilon_{\mathrm{inj}}=1.62\times10^{-5}$
in close vicinity of the shock and at some distance in the foreshock.
One can see, first of all, that the spectra resulting from the CSA
simulation are more intense at lower wavenumbers than the SOLPACS
spectra. Moreover, the wave generation in the CSA simulation reaches
smaller wavenumbers than in the case of the SOLPACS one. This is in
agreement with the difference in the cut-off energy $E_{\mathrm{c}}$
in the corresponding particle energy spectra at the shock. Also, in the CSA spectra, the spectrum amplification drops off at $k\sim10^{-3}\,\mathrm{m^{-1}}$,
whereas in the SOLPACS spectra the amplification extends up to the
highest wavenumbers.

\begin{figure*}
\begin{centering}
\includegraphics[width=\columnwidth]{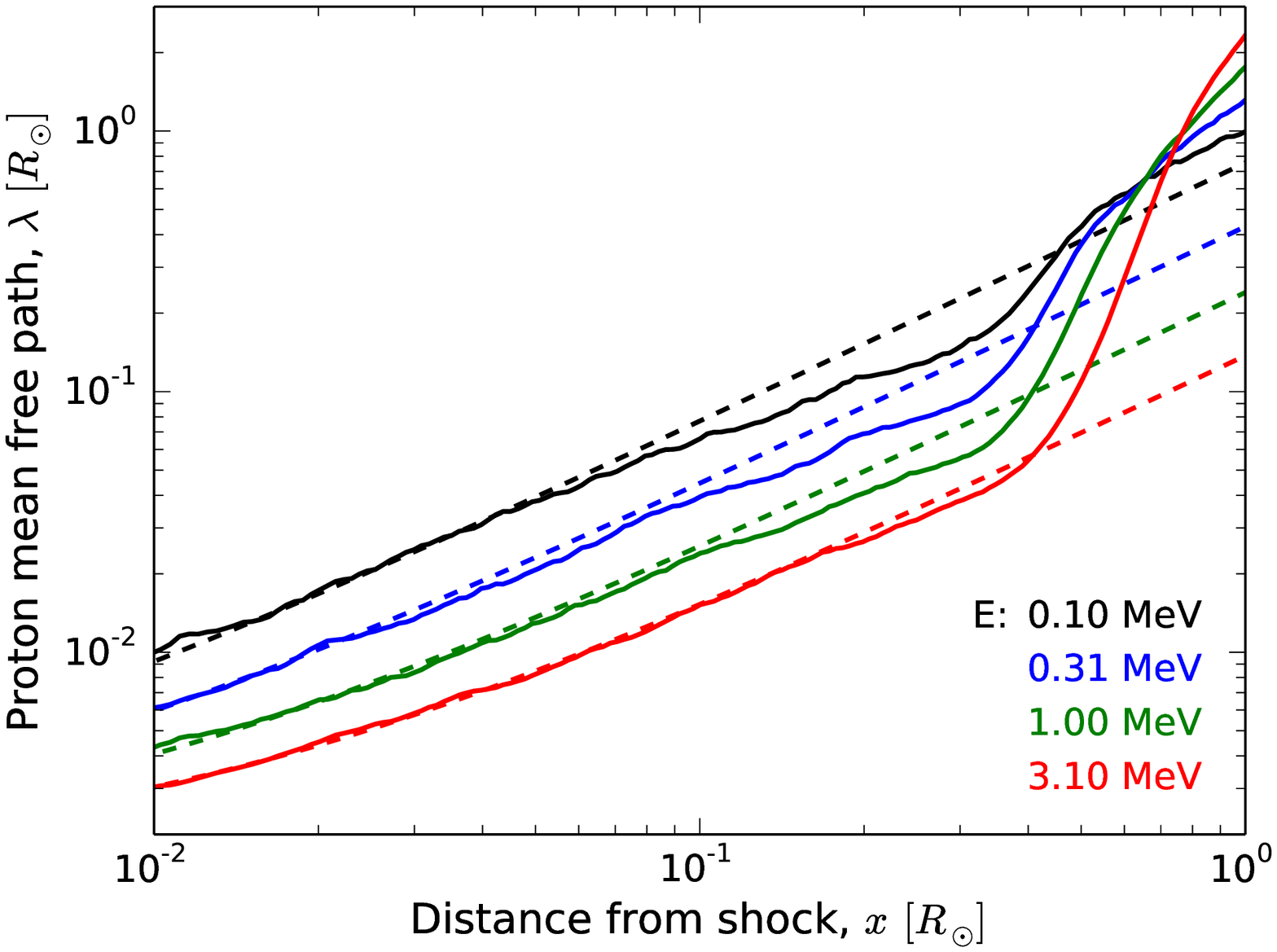}\includegraphics[width=\columnwidth]{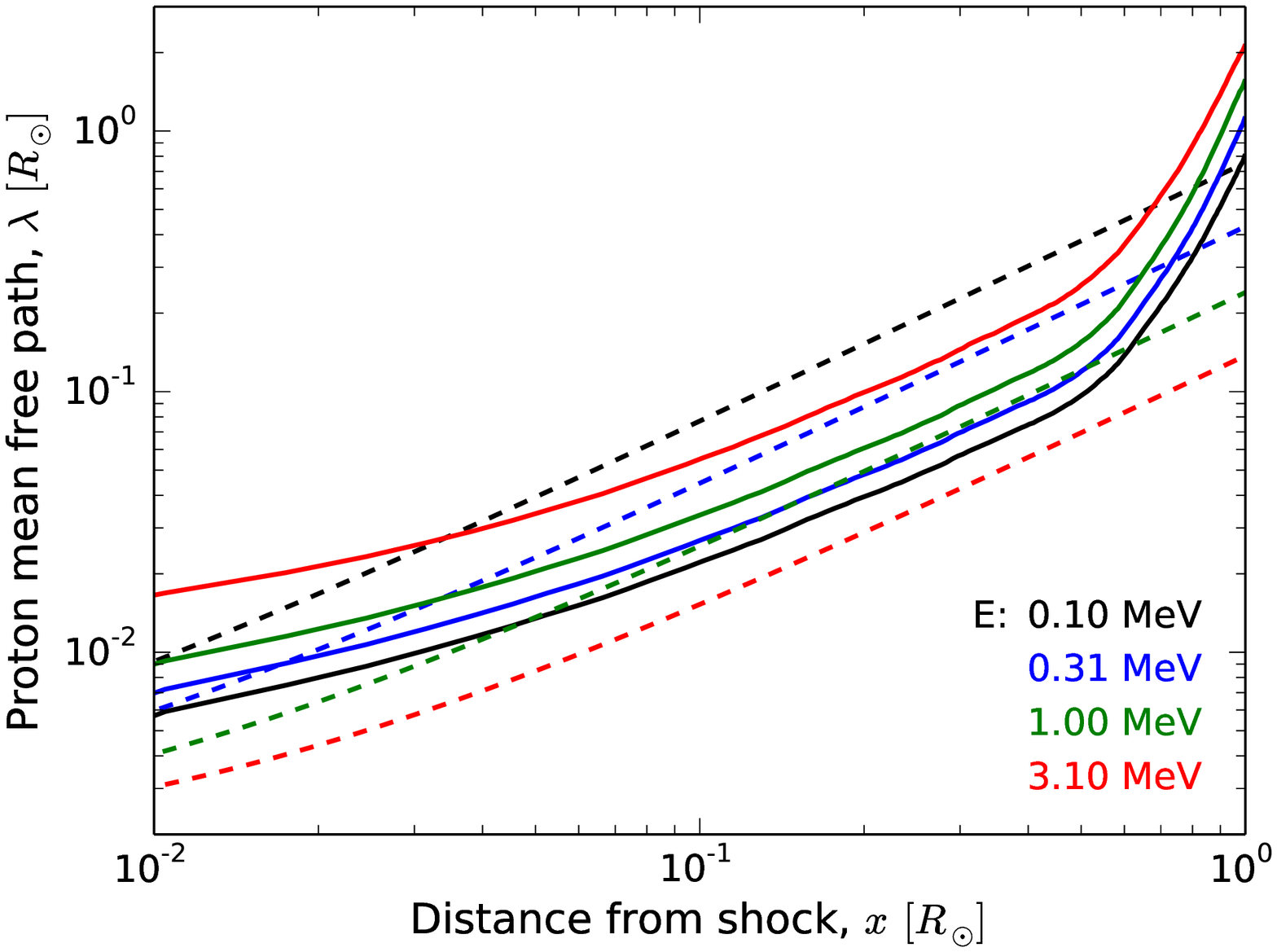}
\par\end{centering}

\protect\caption{\label{fig: mfp_distance }Proton mean free path as a function of
distance from the shock, calculated for different energies for $\epsilon_{\mathrm{inj}}=1.62\times10^{-5}$
at $t=580\,\mathrm{s}$. Left panel: CSA (continuous curves) vs. steady-state
theory (dashed curves); right panel: SOLPACS (continious curves) vs.
steady-state theory (dashed curves).}
\end{figure*}

In Fig. \ref{fig: mfp_distance }, we present spatial distributions
of the particle mean free path in the foreshock at different energies,
calculated from CSA and SOLPACS data along with the mean free paths
predicted by Bell's steady-state theory. To obtain the theoretical
mean free path, the same value of the correction factor $\alpha=1.8$
is applied. The mean free path resulting from CSA was calculated as
$\lambda=\upsilon/\nu$, which results from Eq. (\ref{eq:mfp_definition})
supplemented by Eq. (\ref{eq: Isotrop diff coef}), and the mean free path resulting
from SOLPACS was calculated with Eq. (\ref{eq:mfp-numerical calculation}),
in which we took the resonance gap filling into account. One can
see that the mean free path calculated based on the SOLPACS simulation
data increases with energy, whereas the mean free path calculated using
the steady-state theory demonstrates the opposite dependence on energy.
The mean free path resulting from the CSA simulation data has a more
complex behaviour, although this mean free path demonstrates similarity with the steady-state
theory prediction close to the shock. 

\begin{figure*}
\begin{centering}
\includegraphics[width=\columnwidth]{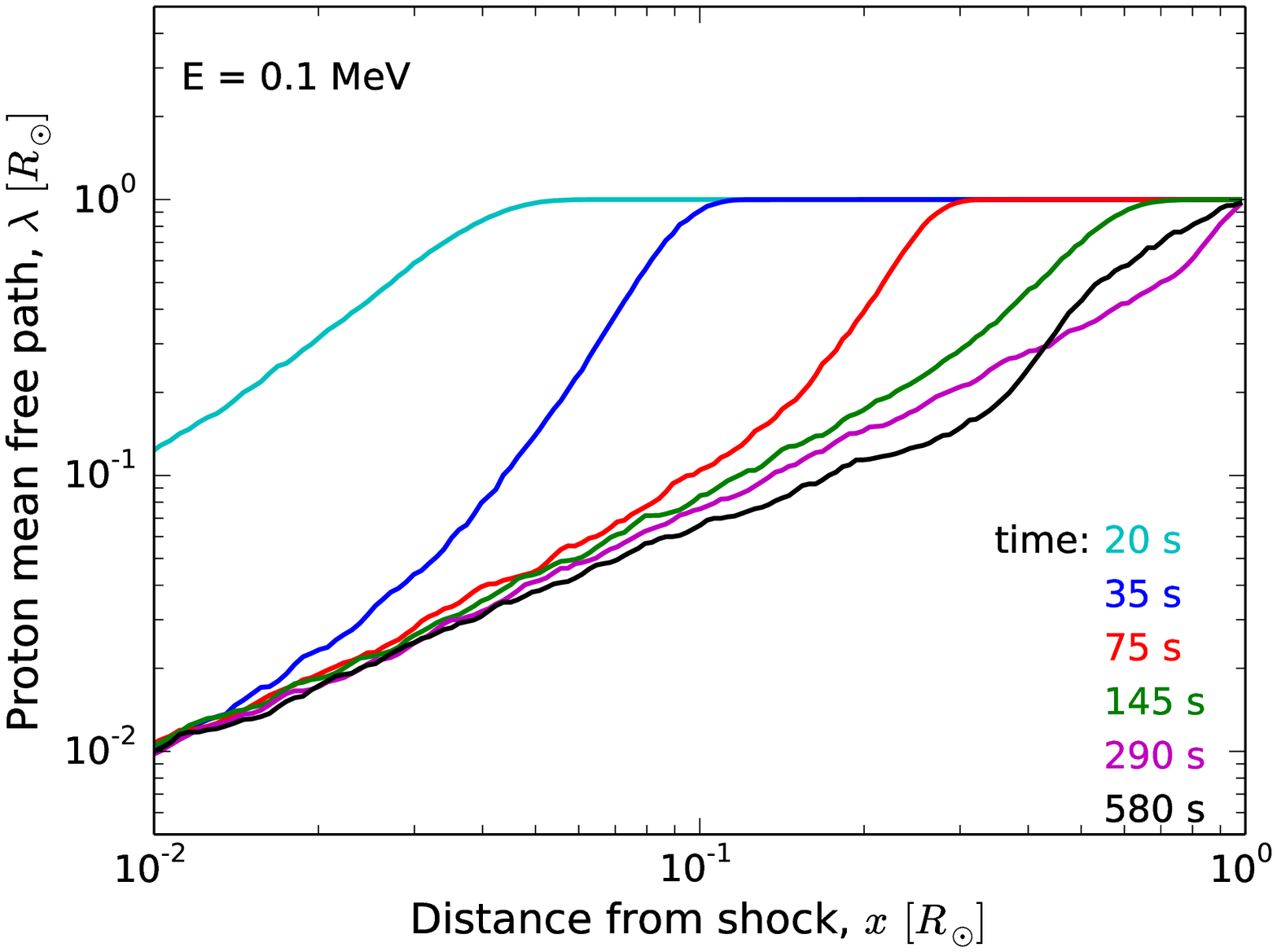}\includegraphics[width=\columnwidth]{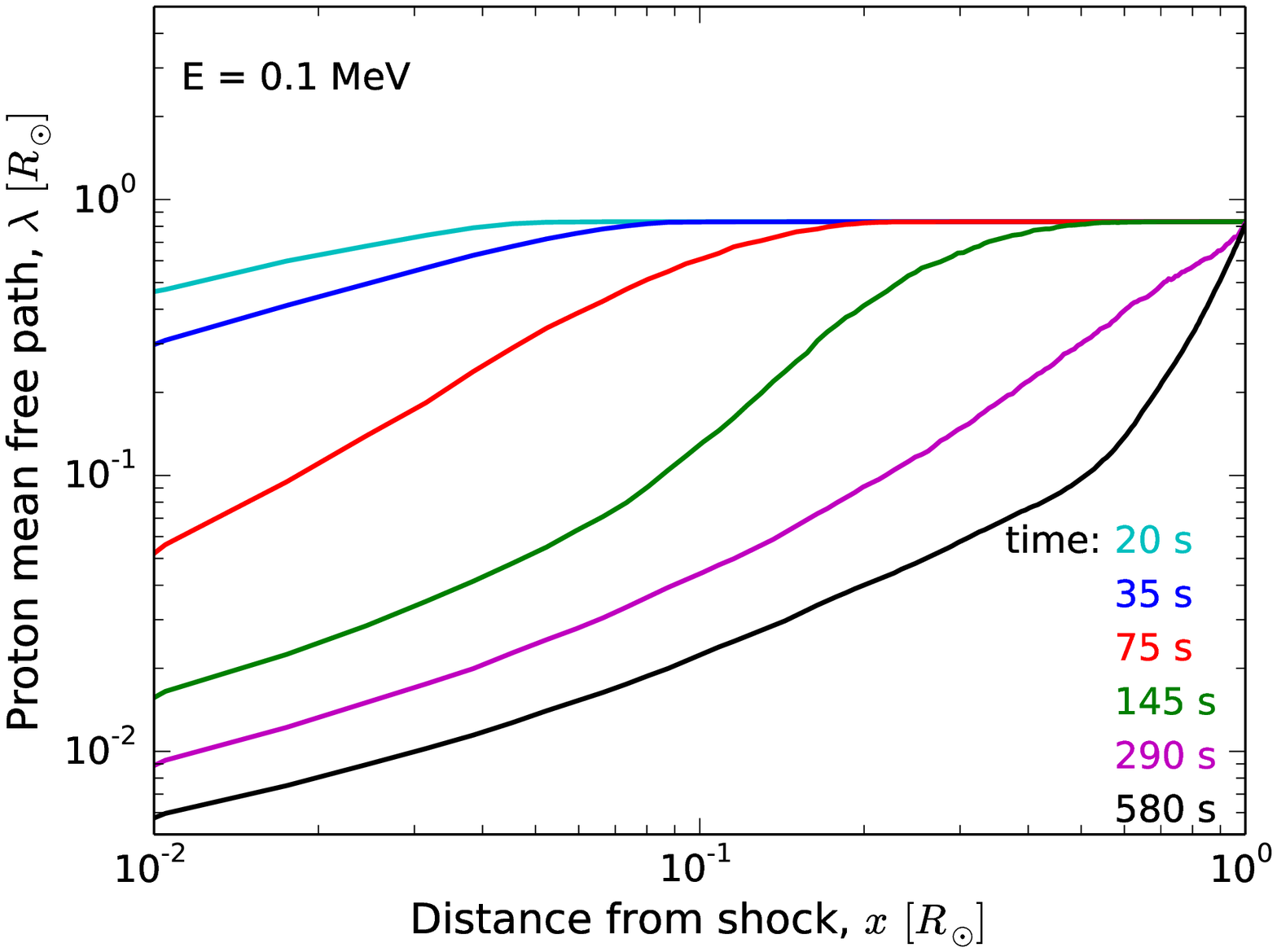}
\par\end{centering}

\protect\caption{Time evolution of the proton mean free path in the foreshock at $E=0.1\,\mathrm{MeV}$
in CSA (left panel) and SOLPACS (right panel) for $\epsilon_{\mathrm{inj}}=1.62\times10^{-5}$.\label{fig:mfp-time-evolution}}
\end{figure*}

\begin{figure}
\begin{centering}
\includegraphics[width=\columnwidth]{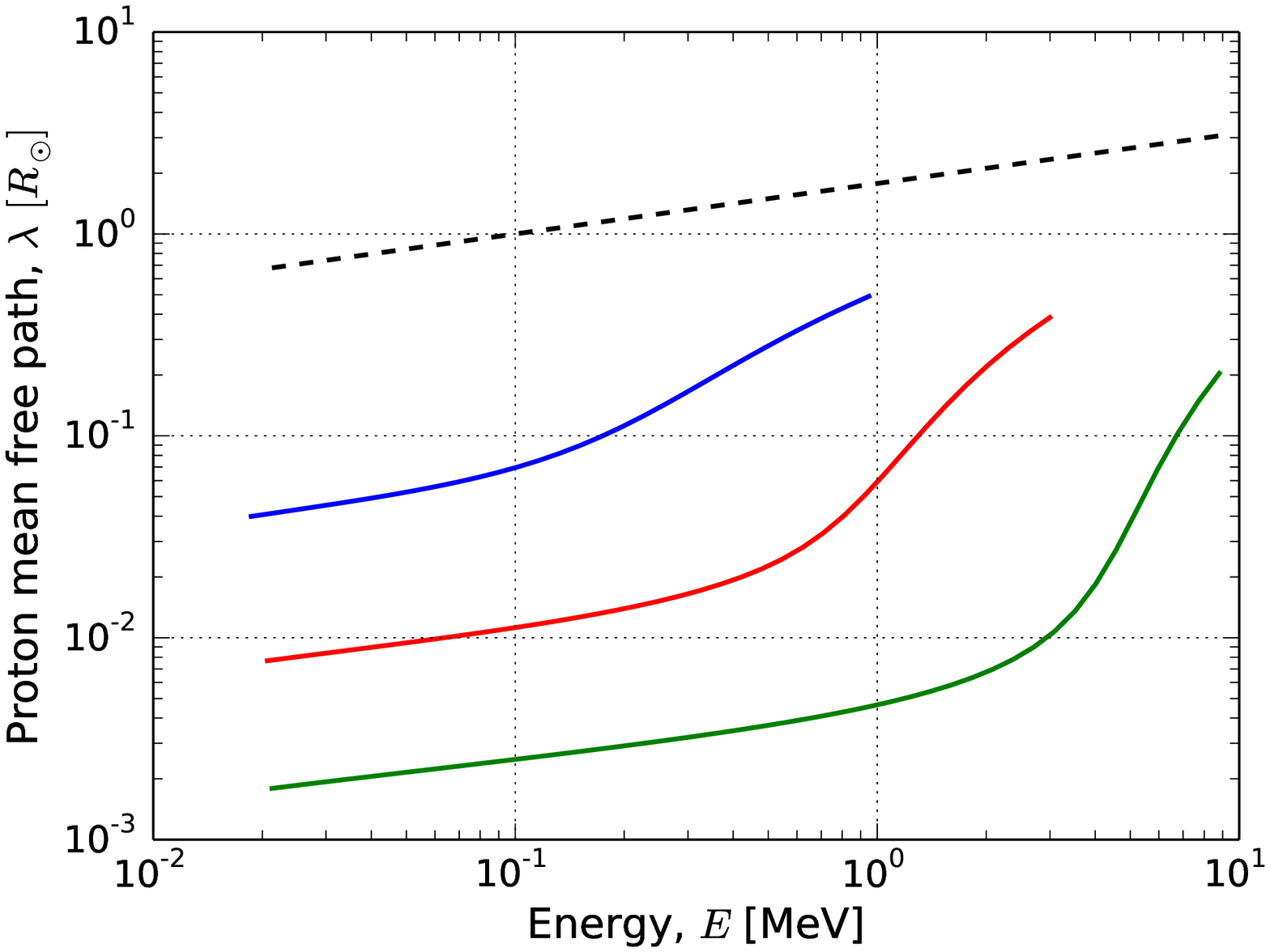}
\par\end{centering}

\protect\caption{Proton mean free path as a function of energy in the vicinity of the
shock obtained with SOLPACS for different values of the injection
parameter: $\epsilon_{\mathrm{inj}}=1.62\times10^{-6}$ (blue), $\epsilon_{\mathrm{inj}}=5.40\times10^{-6}$
(red) and $\epsilon_{\mathrm{inj}}=1.62\times10^{-5}$ (green) at $t=580\,\mathrm{s}$. Dashed line corresponds to the mean free path in the background plasma
with the wave spectrum $I_{\mathrm{w}}\propto k^{-3/2}$ ($\lambda\propto E^{1/4}$).
\label{fig:mfp_energy}}
\end{figure}

Figure \ref{fig:mfp-time-evolution} demonstrates the evolution of
the foreshock mean free path with time in CSA and SOLPACS. In CSA
the mean free path achieves a steady state but in SOLPACS it does
not. The behaviour of the mean free path at higher energies is qualitatively
similar to the presented case for $E=0.1\,\mathrm{MeV}$. Figure \ref{fig:mfp_energy}
shows the SOLPACS energy dependence of the mean free path in the vicinity
of the shock.

\section{Discussion}

\begin{figure*}
\begin{centering}
\includegraphics[width=\columnwidth]{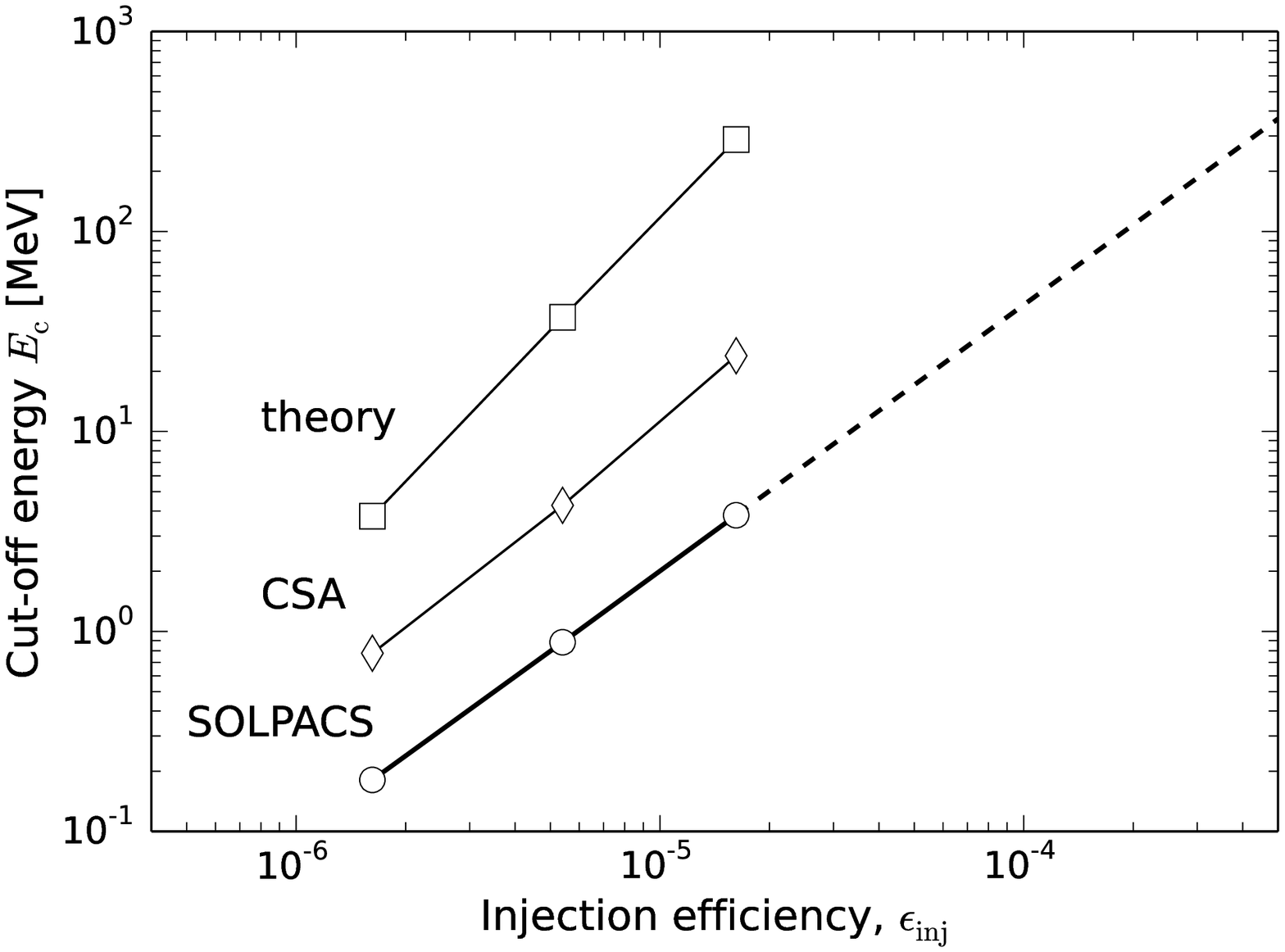} \includegraphics[width=\columnwidth]{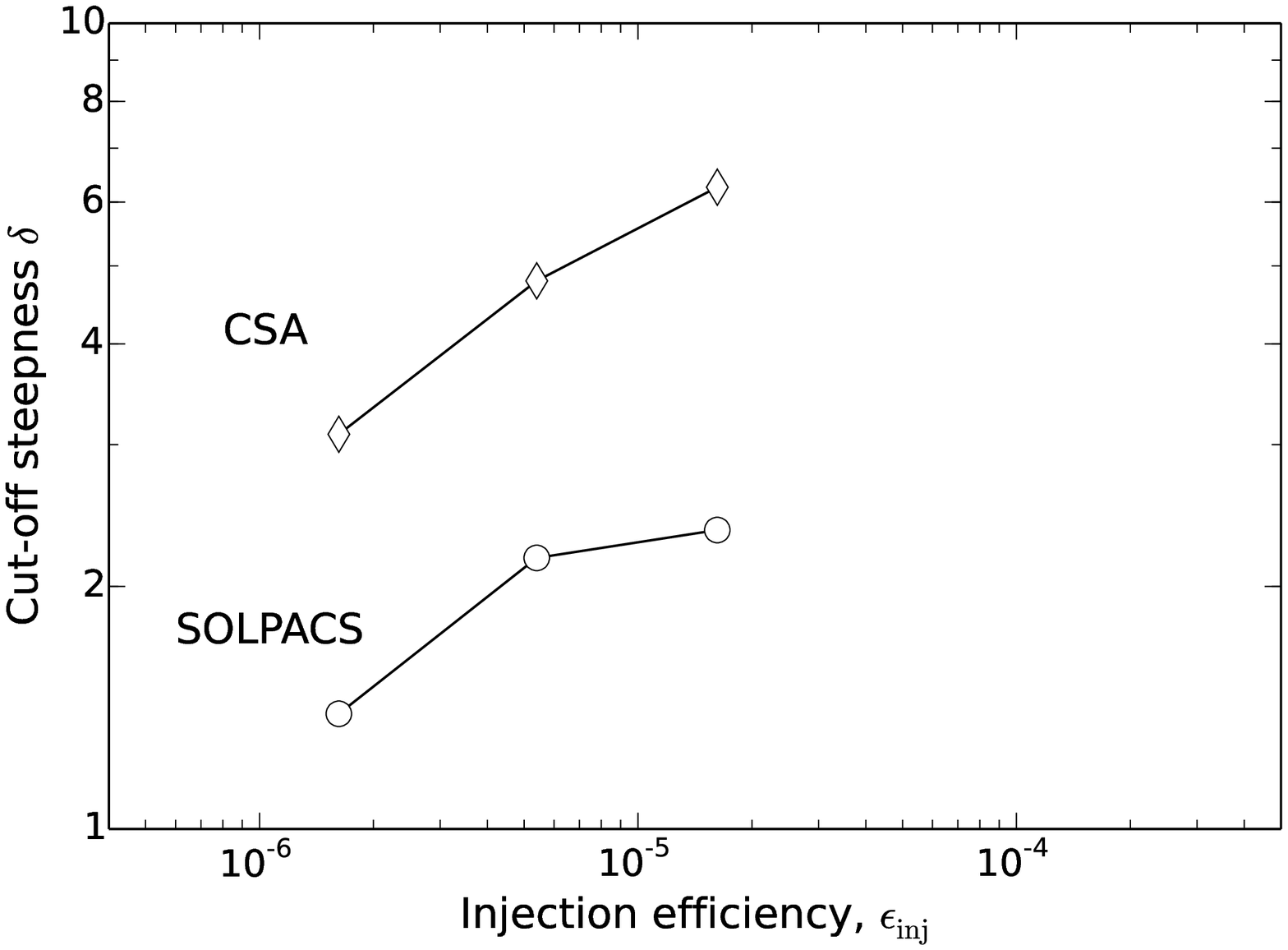}
\par\end{centering}

\protect\caption{\label{fig:Cut-off energy}Left panel: cut-off energy $E_{\mathrm{c}}$
versus injection efficiency $\epsilon_{\mathrm{inj}}$, resulting
from the simulations and steady-state theory. Dashed curve represents
a power-law extrapolation of SOLPACS results to larger values of $\epsilon_{\mathrm{inj}}$.
Right panel: cut-off steepness $\delta$ versus injection efficiency
$\epsilon_{\mathrm{inj}}$, resulting from the simulations.}
\end{figure*}

The simulations performed demonstrate that anisotropic pitch-angle
scattering and the corresponding modification in the wave growth rate
lead to less efficient acceleration of particles at a shock. We can
compare the acceleration efficiency provided by different models by
comparing spectral cut-off energies (Fig. \ref{fig:Cut-off energy},
left panel). The theoretical values of the cut-off energy were calculated
using Eq. (\ref{eq: Cut-off momentum}). One can see that the anisotropic
pitch-angle scattering provides an additional downscaling of the cut-off
energy (approximately by a factor of 5) over the CSA result, which
is already an order of magnitude lower than that provided by Bell's
steady-state theory. The SOLPACS cut-off energies are well fitted
by a power law that one can extrapolate to larger (but still reasonable)
values of the injection efficiency. One can see that for $\epsilon_{\mathrm{inj}}=5\times10^{-4}$
our extrapolation yields $E_{\mathrm{c}}\approx300\,\mathrm{MeV}$.
So, particle acceleration with SOLPACS is still efficient in the sense
that particles can be accelerated to relativistic energies within
ten minutes or less. On the right panel of Fig. \ref{fig:Cut-off energy},
we present the values of the cut-off steepness $\delta$ obtained
in the simulations. We see that SOLPACS provides less abrupt
spectral cut-offs compared to CSA, although the spectral form is still
somewhat steeper than in observations of large SEP events (see e.g.
\citealt{TylkaDietrich2009}; \citealt{AfanasievVainioKocharov-2014}).
However, it is difficult to make a reliable prediction for $\delta$
at large $\epsilon_{\mathrm{inj}}$, based on the data available from
the present simulations. This problem will be addressed in forthcoming
work.

The error analysis of the values of the fitted spectral parameters
(see Table \ref{tab:Best-fit-parameters}) shows that the power-law
index $\beta$ of the particle spectrum obtained with SOLPACS for
$\epsilon_{\mathrm{inj}}=1.62\times10^{-5}$ is  about 8\% lower
than the theoretical index. We speculate that this happens because of ballistic
removal of low-energy particles from the shock since it should take a
longer time for these sorts of particles to scatter over $\mu=0$ than for higher-energy
particles (the resonance gap is wider for lower-energy particles than for
higher-energy particles). The existence of the plateau-like feature in
the spatial distributions of low-energy particles and its absence
for high-energy particles (Fig. \ref{fig:Time evolution of spatial distribs})
supports this idea. Therefore, the plateau feature represents low-energy
particles that propagated to those distances early in the simulation,
i.e. when their turbulent trapping was not very efficient. At later
times, the trapping becomes efficient and does not allow any new low-energy
particles to reach those distances, which is consistent with the large
intensity gradient of low-energy protons close to the shock.

The distinctions between the wave spectra resulting from CSA and SOLPACS,
shown in Fig. \ref{fig:Simulated wave spectra}, are  due to different
forms of the resonance condition of pitch-angle scattering employed
in the codes. The top row of Fig. \ref{fig:Simulated wave spectra}
shows  wave spectra at distance $x=3.5\times10^{-3}\, R_{\odot}$
from the shock. Using Eq. (\ref{eq: SST x0}) of Bell's steady-state
theory, along with the corresponding resonance condition $p=m_{\mathrm{p}}\Omega_0/k$,
we estimate that $k_{\mathrm{b}}=4.20\times10^{-4}\,\mathrm{m^{-1}}$.
For the bottom row of plots in Fig. \ref{fig:Simulated wave spectra}, corresponding to
$x=0.13\, R_{\odot}$, we find $k_{\mathrm{b}}=2.26\times10^{-5}\,\mathrm{m^{-1}}$.
For $\sigma=4$, Bell's theory predicts $x_{0}\propto k^{-1}$ and
$I_{\mathrm{w}}(k)\propto k^{-2}$ at $k\ll k_{\mathrm{b}}$ and $I_{\mathrm{w}}(k)\propto k^{-3}$
at $k\gg k_{\mathrm{b}}$. One can see that the SOLPACS wave spectra have
variable spectral indices $q\lesssim2$, which
thus contradicts Bell's steady-state theory. By contrast, the CSA spectra
(at the wavenumber interval of $10^{-4}\lesssim k\lesssim10^{-3}\,\mathrm{m^{-1}}$,
where the bulk of particles is in resonance) have spectral indices $q$
between 2 and 3, and become steeper with distance, in agreement with
the steady-state theory. The asymptotic form $\sim k^{-2}$
of the SOLPACS spectra is consistent with earlier studies of
wave generation by anisotropic particle distributions, using the full
resonance condition (\citealt{LeeIp-1987}; \citealt{GordonLeeMobius-1999};
\citealt{SchlickeiserVainioBottcher-2002}; \citealt{AfanasievVainio-2013}).

The shape of the wave spectrum governs the energy dependence of the
particle mean free path in the foreshock (Fig. \ref{fig: mfp_distance }).
For $\sigma=4$, the steady-state theory (Eq. \ref{eq:SST mfp}) gives
$\lambda\propto E^{-1/2}$ (non-relativistic regime) at large distances
from the shock ($x\gg x_{0}$). When approaching the shock the differences
between mean free paths corresponding to different energies reduce
and vanish at the shock. In the case of CSA simulations, $|k_{\mathrm{res}}|=\Omega/\upsilon$,
$I_{\mathrm{w,res}}\propto|k_{\mathrm{res}}|^{-q}$. Therefore, $\lambda=\upsilon/\nu\propto E^{-\left(q/2-1\right)}$.
Thus, as the self-generated wave spectrum in CSA has  spectral
indices $q>2$ at the resonant wavenumbers, the mean free path decreases
with energy, as in the steady-state theory. In the case of SOLPACS
simulations, the wave spectrum is characterized by $q\apprle2$ and
the mean free path increases with energy (Fig. \ref{fig:mfp_energy}),
in contrast to CSA and steady-state theory.

Figures \ref{fig:Time evolution of spatial distribs} and \ref{fig:mfp-time-evolution}
clearly demonstrate the effect of high-energy particles on the scattering
conditions of low-energy particles in the foreshock, which is missing from
 steady-state theory and CSA, but captured by SOLPACS. The streaming
high-energy particles constantly generate Alfv\'en waves that interact
with low-energy particles, enhancing their pitch-angle scattering
and facilitating their return to the shock. This process is partly counter-balanced
by the influence of the resonance gap in the $\mu$-space, but in SOLPACS
the effect decreases with time because of increasingly efficient filling
of the resonance gap. The persistent increase of the wave energy at
wavenumbers resonant with low-energy particles leads to the reduction
of the particle mean free path close to the shock during the whole
simulation. This is in contrast to the picture provided by the steady-state
theory and CSA, in which particles are bound to particular wavenumbers
(due to the simplified resonance condition used) and the mean free
path reaches the steady state together with the particle distribution
at the resonant energy. The continuous reduction of the mean free
path in SOLPACS due to continuous acceleration of ions to higher and
higher energies leads to a long-lasting decrease in the particle intensity
in the foreshock at low energies. 

Finally, it should be emphasized that our results highlight the need for 
self-consistent simulations of DSA at oblique shocks. There are theoretical
arguments (e.g. \citealt{EllisonBaringJones-1995}, \citealt{GiacaloneJokipii-2006})
suggesting that particle acceleration rate in an oblique shock should
be enhanced compared to the strictly parallel shock case. On the other
hand, CSA simulations of oblique shocks indicate reduced Alfv\'en
wave generation due to diminished injection efficiency compared to
the strictly parallel shock case (\citealt{BattarbeeVainioLaitinen-2013}).
Therefore a competition between these factors should be expected and
it will be particularly interesting to see what will be the effect of using the
full resonance condition. This problem will be addressed quantitatively
in a future paper.

\section{Summary and conclusions}

In this study, we have presented Monte Carlo simulations of DSA of
protons interacting self-consistently with Alfv\'en waves in the
upstream region of a parallel shock. The simulations were carried
out using the CSA code employing a simplified resonance condition
and, thus, isotropic pitch-angle scattering of particles. We also employed the
SOLPACS code using the full resonance condition and, thus, taking
 anisotropy of pitch-angle scattering into account. We  compared
the simulation results provided by the two codes with each other and
with the steady-state theory. 

In all the simulations, the resulting forms of the particle energy
spectrum at the shock are qualitatively similar, since they are given by a
power-law with a steep super-exponential cut-off. However, anisotropic
pitch-angle scattering and the associated reduced wave growth at low
wavenumbers yield a lower (about 5 times) cut-off energy than the
cut-off energy obtained under isotropic scattering and stronger wave
growth at low wavenumbers. The latter cut-off energy is an order of
magnitude lower than that provided by steady-state theory.

The spatial distributions of particles and waves in the foreshock
provided by CSA after ten minutes are in agreement with Bell's steady-state
theory predictions. Correspondingly, the particle mean free path calculated
from CSA data has the same dependence on particle energy as the theoretical
mean free path. The spatial distributions of particles obtained from
SOLPACS data reveal a much more complicated picture than predicted
by Bell's steady-state theory. The general result arising from the
SOLPACS simulations is that there is a lack of lower-energy particles
and an excess of higher-energy particles in the foreshock (except
in a region very close to the shock) compared to the theoretical
prediction. 

The SOLPACS mean free path has the opposite energy dependence than
that obtained in the CSA simulations and predicted by the theory. The low-energy
mean free path produced by CSA reaches a steady state within the simulation
time, whereas that produced by SOLPACS does not.

Based on the analysis of the results obtained, we conclude that particle
acceleration with anisotropic pitch-angle scattering should be able
to produce relativistic particles for moderate particle injection
rates ($\epsilon_{\mathrm{inj}}<10^{-3}$) in less than ten minutes,
in accordance with observations of relativistic SEP events. However,
reduction in the particle acceleration efficiency due to anisotropic
pitch-angle scattering and reduced wave growth needs to be taken into
account when estimating the maximum particle energy obtained in a
given acceleration time. Finally, the evolution of the foreshock is
substantially influenced by the effect of high-energy particles on
the  pitch-angle scattering conditions of low-energy particles. This is
not accounted for either in Bell's theory or in simulations employing
the simplified resonance condition $k=\Omega/\upsilon$. 
\begin{acknowledgements}
This project has received funding from the European Union\textquoteright s
Horizon 2020 research and innovation programme under grant agreement
No 637324 and from the Academy of Finland (projects 258963 and 267186).
The calculations were performed using the Finnish Grid Infrastructure
(FGI) project (Turku, Finland).
\end{acknowledgements}


\section{Appendix A: Particle energy spectrum fitting}

To fit the particle spectral data obtained from the simulations we
have developed a two-step algorithm based on least-squares fitting
procedure. In the first step, we apply the technique by Vainio et
al. (2014), which consists in fitting the power-law part of the spectrum
separately from its steep tail part, and obtain preliminary values
of the best-fit parameters (Eq. \ref{eq:spectrum fitting function}).
These values are then used as the initial guess values for the precise
fitting in the second step. Furthermore, in the second step we use
information about errors. The fitting itself is performed using information
about the statistical error in the data and the error due to binning
of the data, but the fitting parameter error estimations given in
Table \ref{tab:Best-fit-parameters} also account for the systematic
error due to the non-power-law form of the spectrum at low energies. 

The statistical error is calculated based on the Poisson statistics,
in which the standard deviation $\sigma_{\mathrm{P}}\sim\sqrt{N}$,
where $N$ is the number of events. Therefore, as the particle intensity
in a given energy bin is calculated as $I=C\sum_{j=1}^{N}w_{j}$,
then the standard deviation of the particle intensity due to the statistical
error is 
\[
\Delta I_{\mathrm{stat}}=C\left(\sum_{j=1}^{N}w_{j}^{2}\right)^{1/2},
\]
where $C$ is the normalizing coefficient, $w_{j}$ is the weight
of the $j$-th particle, and $N$ is the number of Monte Carlo particles
in the bin (in the case of equal particle weights $w$, we obtain
$\Delta I_{\mathrm{stat}}=Cw\sqrt{N}$ as per Poisson statistics). 

The error due to binning stems from the fact that we fit histograms
that we obtain by distributing Monte Carlo particles by their energy
into different energy bins. This error for a given energy bin can
be calculated from the following:
\[
\Delta I_{\mathrm{bin}}=I(E)-\frac{1}{\Delta E}\intop_{E-\frac{\Delta E}{2}}^{E+\frac{\Delta E}{2}}I(E')dE'.
\]
 Expanding the integrand into the Taylor series up to second order
and assuming that $I(E)\propto E^{-\beta}$, one can obtain
\[
\Delta I_{\mathrm{bin}}=\frac{\beta(\beta+1)}{24}\left(\frac{\Delta E}{E}\right)^{2}I(E).
\]

Both errors can be combined to give
\[
\Delta I=(\Delta I_{\mathrm{stat}}^{2}+\Delta I_{\mathrm{bin}}^{2})^{1/2},
\]
which was used in fitting the particle spectral data. For the CSA
data, we  estimated the statistical error is about an order of
magnitude smaller than the error by binning, whereas for the SOLPACS
data the statistical error  prevails. 

For a given parameter $p$ of the fitting function, the fitting procedure
(at energies $E>E_{0}$) gives the best-fit value, $p_{0,\mathrm{best}}$
and the standard deviation error, $\sigma_{p,\mathrm{fit}}$. To take
into account the systematic error due to the non-power-law form of the
spectrum at low energies, we perform two additional fitting sessions
by starting to fit from the energy $E_{0}-\Delta E$ and from $E_{0}+\Delta E$
correspondingly, and calculate $\Delta p=|p_{\mathrm{best}}-p_{0,\mathrm{best}}|$
for both additional fits. We take the largest $\Delta p_{\mathrm{}}$
and estimate the total error as 

\[
\sigma_{p}=(\sigma_{p,\mathrm{fit}}^{2}+\Delta p^{2})^{1/2},
\]
 which is the value given in Table \ref{tab:Best-fit-parameters}.

\bibliographystyle{aa}
\bibliography{paper_references}

\end{document}